\documentclass[fleqn,usenatbib]{mnras}

\usepackage{newtxtext,newtxmath}

\usepackage[T1]{fontenc}

\DeclareRobustCommand{\VAN}[3]{#2}
\let\VANthebibliography\thebibliography
\def\thebibliography{\DeclareRobustCommand{\VAN}[3]{##3}\VANthebibliography}

\usepackage{graphicx}
\usepackage{amsmath}
\usepackage{subcaption}
\usepackage{xspace}

\graphicspath{ {./figures/} }
\def\reff@jnl#1{{\rm#1\/}}
\def\aj{\reff@jnl{AJ}}                  
\def\araa{\reff@jnl{ARA\&A}}            
\def\apj{\reff@jnl{ApJ}}                
\def\apjl{\reff@jnl{ApJ}}               
\def\apjs{\reff@jnl{ApJS}}              
\def\ao{\reff@jnl{Appl.Optics}}         
\def\apss{\reff@jnl{Ap\&SS}}            
\def\aap{\reff@jnl{A\&A}}               
\def\aapr{\reff@jnl{A\&A~Rev.}}         
\def\aaps{\reff@jnl{A\&AS}}             
\def\baas{\reff@jnl{BAAS}}              
\def\jrasc{\reff@jnl{JRASC}}            
\def\memras{\reff@jnl{MmRAS}}           
\def\mnras{\reff@jnl{MNRAS}}            
\def\pra{\reff@jnl{Phys.Rev.A}}         
\def\prb{\reff@jnl{Phys.Rev.B}}         
\def\prc{\reff@jnl{Phys.Rev.C}}         
\def\prd{\reff@jnl{Phys.Rev.D}}         
\def\prl{\reff@jnl{Phys.Rev.Lett}}      
\def\procspie{\reff@jnl{Society of Photo-Optical Instrumentation Engineers (SPIE) Conference Series}}      
\def\pasp{\reff@jnl{PASP}}              
\def\pasj{\reff@jnl{PASJ}}              
\def\qjras{\reff@jnl{QJRAS}}            
\def\skytel{\reff@jnl{S\&T}}            
\def\solphys{\reff@jnl{Solar~Phys.}}    
\def\sovast{\reff@jnl{Soviet~Ast.}}     
\def\ssr{\reff@jnl{Space~Sci.Rev.}}     
\def\zap{\reff@jnl{ZAp}}                
\def\nat{\reff@jnl{Nature}}             
\def\nar{\reff@jnl{NewA Rev.}}  
\def\na{\reff@jnl{NewA}}  

\newcommand{\ie}{\textit{i.e.}\xspace}
\newcommand{\fref}{Fig.~\ref}
\newcommand{\Fref}{Fig.~\ref}
\newcommand{\eref}{Eq.~\ref}

\newcommand{\tref}{Tab.~\ref}

\newcommand{\sref}{Section~\ref}

\title[Focal Plane Wavefront Sensing with Convolutional Neural Networks]{Focal Plane Wavefront Sensing using Machine Learning: \\
Performance of Convolutional Neural Networks compared to Fundamental Limits}

\author[G. Orban de Xivry et al.]{
G. Orban de Xivry$^{1}$\thanks{E-mail: gorban@uliege.be}
M. Quesnel,$^{1,2}$
P.-O. Vanberg,$^{1,2}$
O. Absil,$^{1}$
G. Louppe$^2$
\\
$^{1}$Space sciences, Technologies and Astrophysics Research (STAR) Institute, Universit\'e de Li\`ege, Belgium\\
$^{2}$Montefiore Institute of Electrical Engineering and Computer Science, Universit\'e de Li\`ege, Belgium\\
}

\date{Accepted 2021 June 02. Received 2021 June 01; in original form 2021 January 24}

\pubyear{2021}

\begin{document}
\label{firstpage}
\pagerange{\pageref{firstpage}--\pageref{lastpage}}
\maketitle

\begin{abstract}

Focal plane wavefront sensing (FPWFS) is appealing for several reasons. Notably, it offers high sensitivity and does not suffer from non-common path aberrations (NCPA).
The price to pay is a high computational burden and the need for diversity to lift any phase ambiguity. If those limitations can be overcome, FPWFS is a great solution for NCPA measurement, a key limitation for high-contrast imaging, and could be used as adaptive optics wavefront sensor. Here, we propose to use deep convolutional neural networks (CNNs) to measure NCPA based on focal plane images. Two CNN architectures are considered:  ResNet-50 and U-Net which are used respectively to estimate Zernike coefficients or directly the phase.
The models are trained on labelled datasets and evaluated at various flux levels and for two spatial frequency contents (20 and 100 Zernike modes).
In these idealized simulations we demonstrate that the CNN-based models reach the photon noise limit in a large range of conditions.
We show, for example, that the root mean squared (rms) wavefront error (WFE) can be reduced to $< \lambda/1500 $ for $2 \times 10^6$ photons in one iteration when estimating 20 Zernike modes.
We also show that CNN-based models are sufficiently robust to varying signal-to-noise ratio, under the presence of higher-order aberrations, and under different amplitudes of aberrations.
Additionally, they display similar to superior performance compared to iterative phase retrieval algorithms.
CNNs therefore represent a compelling way to implement FPWFS, which can leverage the high sensitivity of FPWFS over a broad range of conditions.

\end{abstract}



\section{Introduction}

High-contrast imaging instruments are now routinely used in ground-based astronomy  to explore circumstellar environments and to detect exoplanets. To achieve such a feat, they must reach high contrast at small angular separation and thus rely on a precise control of the wavefront. Extreme adaptive optics (AO) systems correct the corrugated wavefront caused by atmospheric turbulence and provide near-perfect diffraction limited point spread functions, which can then be effectively suppressed by a coronagraph.
However, the contrast, or likewise the exoplanet detectability, may still be limited by non-common path aberrations (NCPA)
between the wavefront sensor arm and the scientific path. These NCPA are quasi-static with minute to hour timescales  due to slowly evolving instrumental aberrations and beam wander related to temperature, humidity, and mechanical changes.
Because of their nature, they appear essentially at low spatial frequencies. These properties make them challenging to remove in post-processing and detrimental to the final contrast.
In this respect, focal plane wavefront sensing (FPWFS) with the scientific detector is an appealing solution.
In addition to getting rid of NCPA and chromatic errors, FPWFS offers high sensitivity that is only surpassed by the Zernike wavefront sensor \citep{Guyon:05}. It is also simple opto-mechanically and necessitates few to no modifications of the optics. This low complexity means less risk of failure and less maintenance. An overview of existing FPWFS techniques for high-constrat imaging instruments can be found in \cite{Jovanovic:18} but the interest for FPWFS goes well beyond NPCA measurement.
Some aberrations are not well measured by pupil WFS such as phase discontinuities caused by the presence of spiders \citep[so called petalling and low wind effect, e.g.][]{Vievard:19}.
Other applications range from co-phasing segmented mirrors \citep[e.g.][]{Delavaquerie:10} to real-time adaptive optics systems  \cite[e.g.][]{Keller:12, Korkiakoski:12}.
However the price to pay  of FPWFS is typically a high computational burden, as the problem is non-linear, and it generally requires a source of diversity to effectively lift any phase ambiguity.

In parallel, machine learning algorithms have been developed and applied to phase retrieval and wavefront sensing, in many different fields including astronomy.
 Neural networks were first used for real-time atmospheric compensation and co-phasing \citep{Angel:90,Sandler:91}, and retrieval of static aberration in the Hubble Space Telescope \citep{Barrett:93}.
These techniques have then been used more broadly in the field of AO, to reduce Shack-Hartmann WFS slope errors \citep{Montera:96}, to perform open-loop AO tomographic reconstruction \citep{Osborn:14}, or to predict wavefront and reduce temporal errors \citep[e.g.][]{McGuire:99,Jorgenson:92,Liu:20}.
The non-linear nature of neural networks makes them good candidates to solve the non-linear phase retrieval problem. Despite these early results, the lack of generalization power and the poor scaling of the networks ultimately limited the achievable performance. Later on, convolutional neural networks were introduced \citep{LeCun:1990,Krizhevsky:2012}.
Specifically designed for images, CNNs use successive convolution operations, learning from group of pixels and assembling progressively more complex patterns.
More recent works have applied such CNNs to non-linear wavefront reconstruction \citep{Swanson:18,Landman:20}, wavefront prediction  \citep{Swanson:18,Swanson:21}, to extend the usable range of Lyot-based low order wavefront sensors \citep{Allan:20:LLOWFS} and of Zernike phase-contrast wavefront sensors \citep{Allan:20:ZWFS}, and to focal plane wavefront sensing \cite[e.g.][]{Paine:18, Andersen:2020}. 

Deep learning techniques are in fact burgeoing in all optical applications using phase retrieval, ranging from biomedical microscopy \cite[e.g.][]{Krishnan:2020, Cumming:2020} to holography \cite[e.g.][]{Peng:2020} and astronomy.
The specific  application to image-based wavefront sensing has been investigated in several recent works that we attempt to summarize in the following.
\cite{Naik:20} used a compact CNN for object-agnostic wavefront sensing, infering up to six Zernike coefficients, but reported a poorly sensed coma.
\cite{Wu:20} trained a CNN for fast inference of 13 Zernike coefficients and obtained mild improvements for input aberrations of around 2 rad rms.
\cite{Nishizaki:19} proposed to extend the design space of wavefront sensor using deep learning where the inputs are preconditioned images such as overexposed, defocussed, or scattered images.
\cite{Guo:19} used a direct phase-map-output CNN model for the inference of up to 64th Zernike mode with input WFE from about 1.5 up to 4.5 rad rms. They obtained residual errors in the range of approximately 0.45 to 0.82 rad and validated their approach experimentally.
\cite{Paine:18} used the Inception v3 architecture to expand the capture range of gradient-based optimization methods. They applied their approach to the JWST aperture and consider up to 18 Zernike coefficients with input WFE in the range of 1.57 up to 25.1 rad rms. The residual WFE after the CNN is on average 2.3 rad rms. The trained CNN  provides here good initial estimates to a second stage gradient-based method.
\cite{Andersen:2019,Andersen:2020} studied the potential of real-time image sharpening using ResNet and Inception v3 models to estimate Zernike coefficients  from pairs of in and out-of-focus images that are blurred by the atmospheric turbulence. They included the effect of noise, guide star magnitude, polychromaticity, and bit depth.
They explored an aberration regime of about 8 to 13 rad rms ($D/r_0 = 12$ and $21$ respectively, with $D$ the telescope diameter and $r_0$ the Fried parameter) and obtained a residual error of about 1.4 and 2 rad rms respectively by correcting for 36 Zernike modes. When increasing the number to 66 modes, they obtained only a marginal improvement.

While those studies demonstrated the validity of the approach, \ie using a CNN-based framework for FPWFS, it is unclear what really limits the performance reported, nor if CNNs can leverage fully the sensitivity of FPWFS and if they can be applied effectively in a lower aberration regime relevant to NCPA measurements.
We first investigated this lower aberration regime in a short report \citep{Vanberg:19} demonstrating its working principle. We then explored in \cite{Quesnel:20} different numerical aspects and applied our framework to vortex coronagraphic imaging.

In the present paper, our focus is to better understand the limitations of such a CNN-based framework for FPWFS in the context of NCPA measurements. More specifically, we study a regime of up to 1 rad rms input WFE and up to 100 Zernike modes. We also deliberately limit the number of simulated effects, such as e.g. noise sources or higher-order disturbances, to systematically explore the achievable performance of CNNs for FPWFS and compare it to the fundamental limit for wavefront sensing.
First, in \sref{sec:method}, we describe our simulation setup, \ie the data simulator and the CNN architectures used for this work.
In \sref{sec:results}, we analyze our CNN models under idealized and degraded conditions and compare them with the expected photon noise limit. We also consider the implication of the sign ambiguity and the pixel sampling. Finally, in \sref{sec:discussion} we  compare the CNN model to iterative phase retrieval and discuss numerical considerations.
Overall, we demonstrate that CNN-based algorithms can efficiently solve the inverse problem posed by focal plane wavefront sensing.
In particular, our framework is shown to be readily applicable for the measurement of NCPA, and we discuss throughout the paper different considerations for laboratory and on-sky applications, and for  a broader usage such as, e.g. adaptive optics.

\section{Methods and simulation setup}\label{sec:method}

\subsection{Focal plane imaging \& dataset generation}\label{sec:dataset}

One of the keys to the success of deep learning is the availability of a large and representative labelled dataset. In this paper, the data consists of a set of numerically simulated, aberrated point spread function (PSF) pairs: in-focus and out-of-focus. The introduction of this phase diversity ensures the uniqueness \citep{Foley:81,Gons:82,Paxman:92} of the solution while being easy to implement in practice, either by introducing defocus on a deformable mirror or by displacing the detector itself.
The well-known sign ambiguity in the absence of diversity is discussed in \sref{sec:diversity}.

\begin{figure}
  \includegraphics[width=\linewidth]{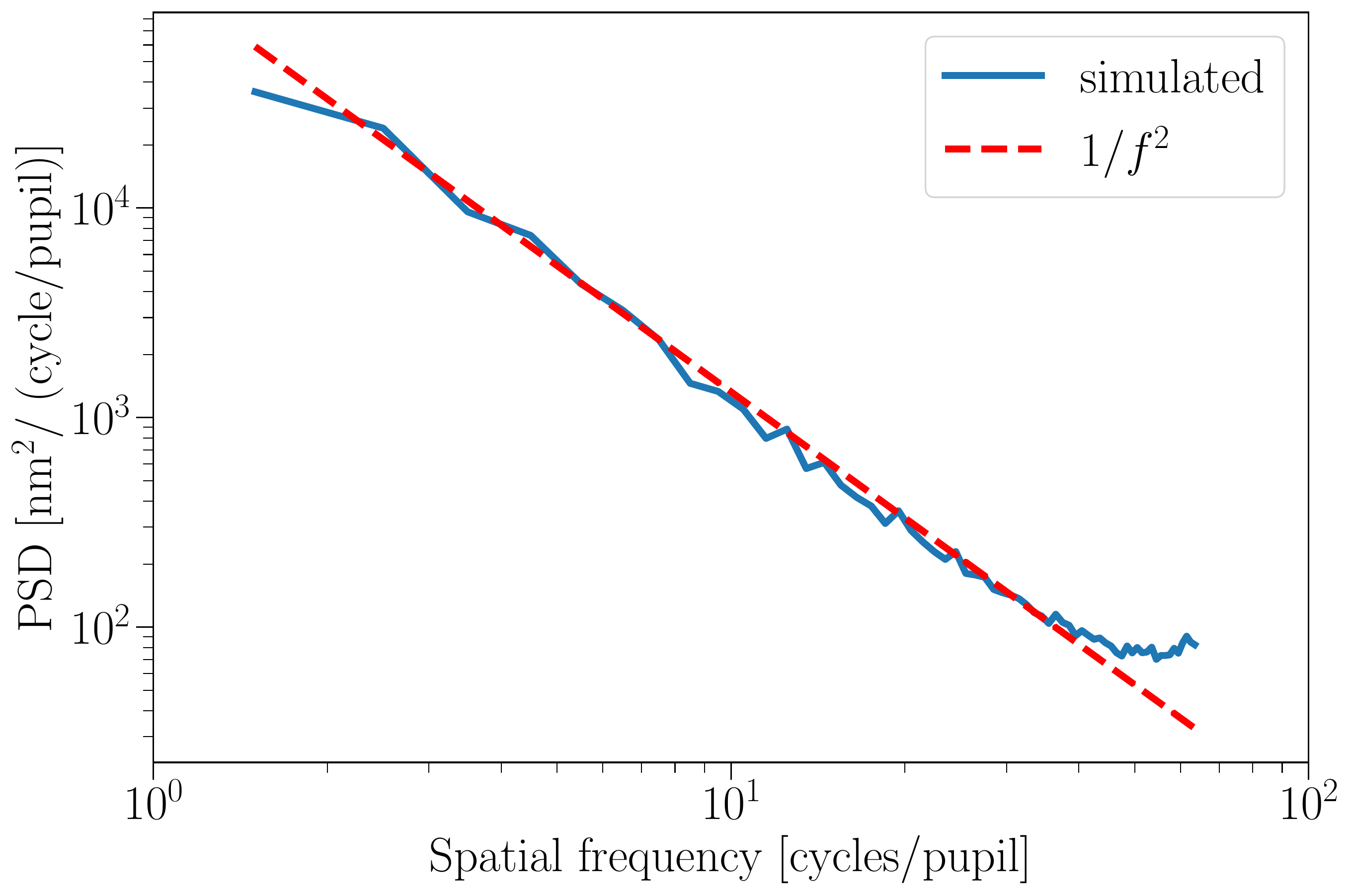}
  \caption{Spatial power spectral density of the generated phase maps reproducing high-quality optical surfaces. }
  \label{fig:dataset_psd}
\end{figure}

\begin{figure}
  \includegraphics[width=\linewidth]{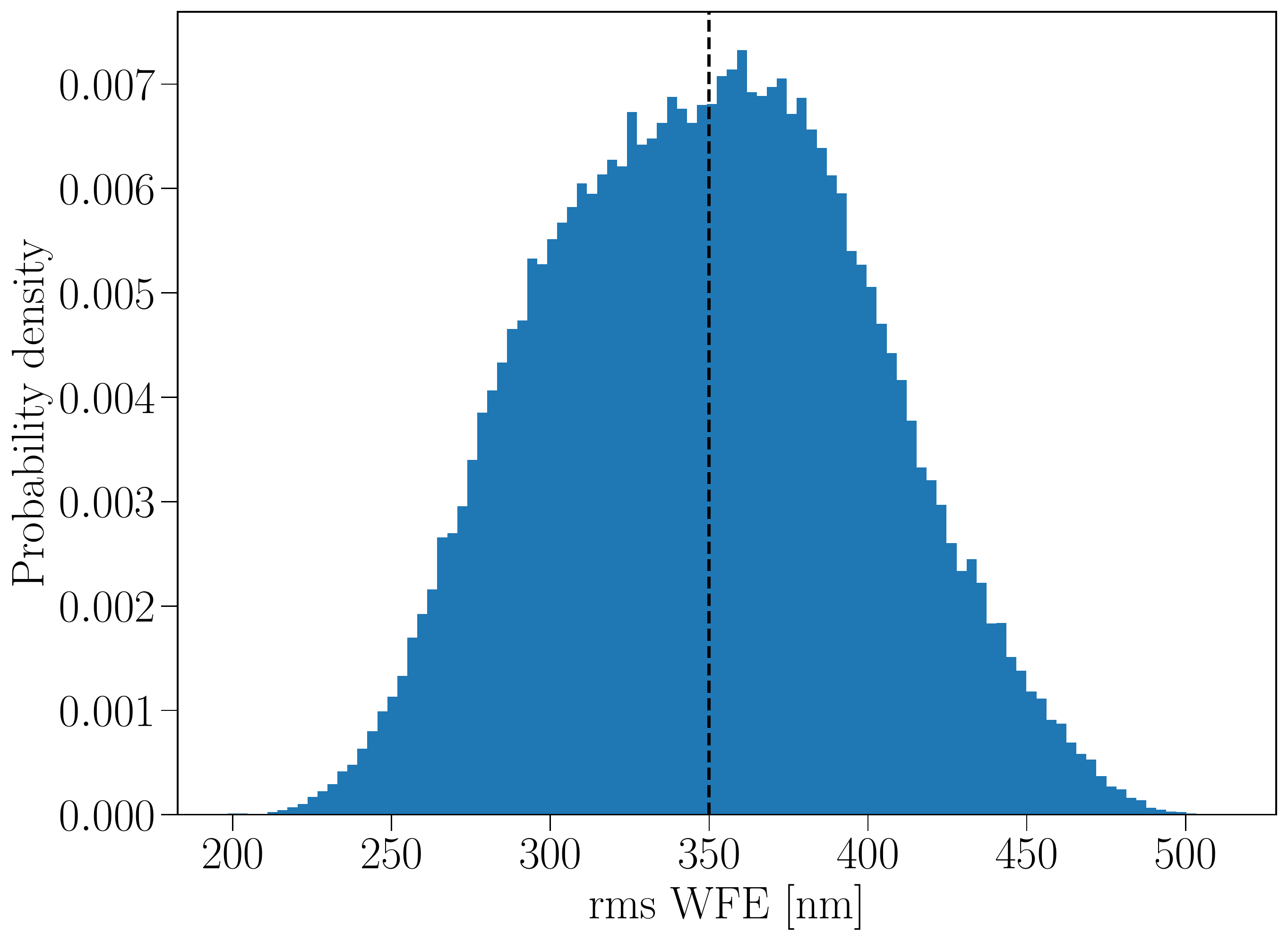}
  \caption{Distribution of the rms WFE of one dataset with a median of 350nm.}
  \label{fig:dataset_hist}
\end{figure}

Since our work is primarily motivated by the measurement and correction of NCPA, we generate phase maps to reflect typical errors of high-quality optical surfaces \citep[e.g.][]{Dohlen:2011}, with a spatial power spectral density (PSD) profile  $S \approx 1/f^2$, where $f$ is the spatial frequency.  The PSD is illustrated in \fref{fig:dataset_psd}.
This is achieved by drawing random Zernike coefficients from a uniform distribution between $[-1, 1]$ and dividing each coefficient by its Zernike radial order.
The coefficients are then scaled to obtain the desired median rms wavefront error (WFE), where the median is calculated over the dataset.
The rms WFE distribution over one of our datasets is illustrated in \fref{fig:dataset_hist}.
This procedure leads to a uniform density distribution for each individual Zernike mode, where the minimum and maximum depend on the Zernike index and are function of the desired median rms WFE and the number of Zernike coefficients.

Finally, the phase maps $ \varphi (x,y)$ are obtained as a linear combination of the Zernike polynomials weighted by the previously generated coefficients. Each one of them is then propagated through the system to produce the corresponding point spread functions $PSF(x,y)$,
\begin{equation}
  PSF(x, y) \propto \left| \mathcal{F}\left[ A(x,y)  \exp{\left(i \varphi(x,y) \right)} \right] \right|^2,
\end{equation}
where $A(x,y)$ is the pupil function. The pupil function considered here is a simple uniformly illuminated circular pupil.
The measurement, \ie the PSF, is finally affected by noise. Here we limit ourselves to photon noise,
and disregard for example detector noises which are technological in nature.
Hence the signal-to-noise ratio of our image is  $\mathrm{SNR}= \sqrt{N_{\rm ph}}$, with $N_{\rm ph}$ the total number of photons in the image.

The image sizes are fixed to 128 $\times$ 128 pixels. The PSFs are sampled by 4.5 pixels over $1\lambda/D$ and the corresponding field-of-view is $\sim 28.5 \lambda / D$. Such PSF sampling can be obtained, for instance, for a wavelength of 2.2$\mu$m, a pixel scale of 0.01\arcsec/pixel and a telescope diameter of 10m. Those parameters are representative of existing instruments such as, e.g., NIRC-2 at the Keck Observatory. Before being saved, the focal plane images are formatted in half-precision (float 16 bits). This step ensures that the theoretical sign ambiguity is perfectly reproduced numerically, \ie that the PSFs generated from phase maps that only differ by the sign of their odd Zernike modes are numerically identical.

\begin{figure}
  \centering
  \includegraphics[width=\linewidth]{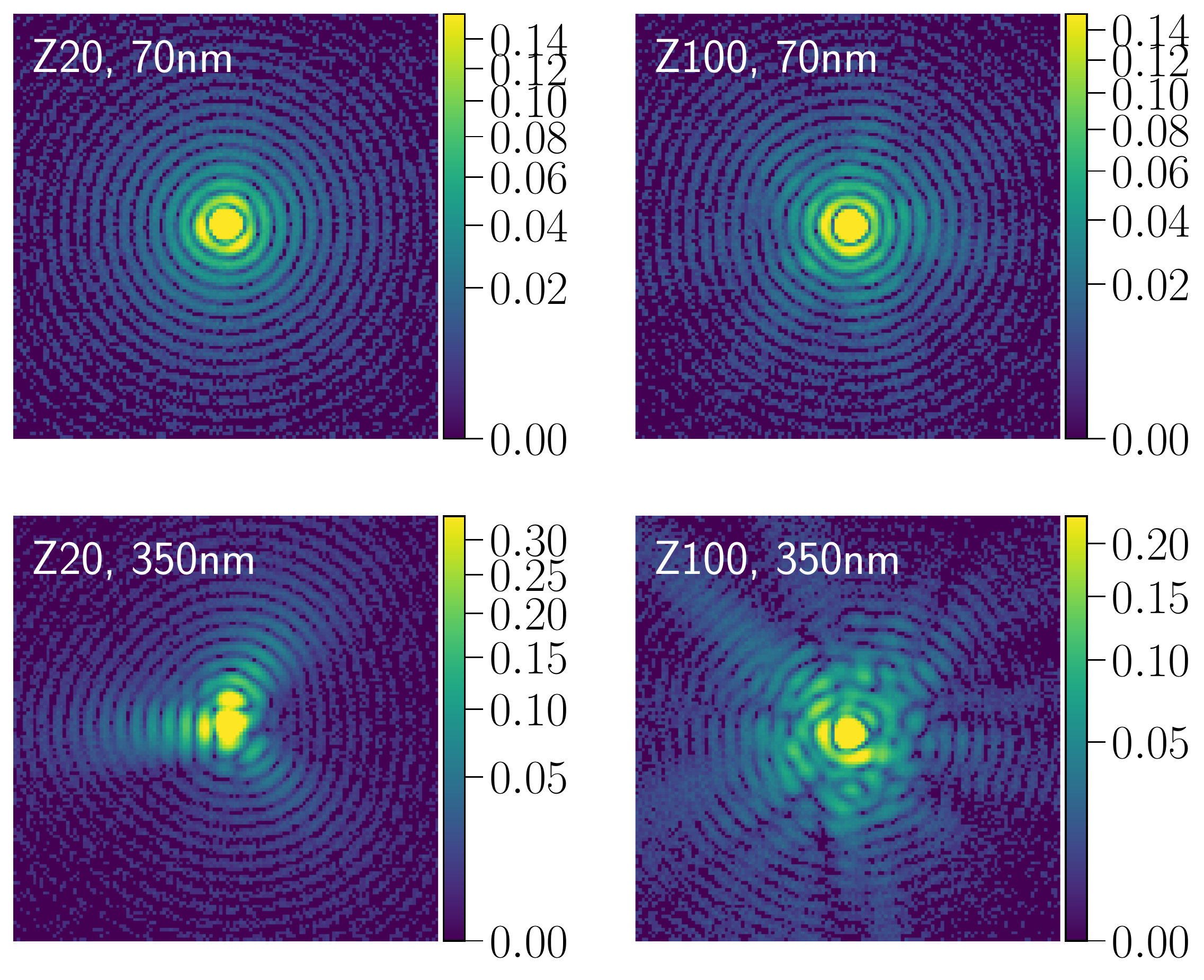}
  \caption{Illustration of the simulated PSFs with a square root stretch and 99\% interval. The signal-to-noise ratio equals 1000. Aberrations distributed over 20  (Left) and 100 (Right) Zernike modes. ``Low'' (Top) and ``high'' (Bottom) aberrations levels. }
  \label{fig:psfs}
\end{figure}

In this work, we consider median rms WFE of 70nm and 350nm at a wavelength of $2.2\mu$m,  thus 0.2 and 1 rad rms respectively. For convenience, we will often refer to those two levels as ``low" and ``high" aberration regime.
We also consider two different numbers of Zernike modes, 20 and 100. We have thus four different scenarii for our following analyses. The resulting PSFs are illustrated in \fref{fig:psfs}.
The introduced phase diversity for the second PSFs is a defocus term set to $\lambda/4$, \ie 550nm rms.
The motivation to limit our training data to those four regimes is twofold. First, the number of modes and the aberration level represent what is typically considered for NCPA correction on 8 to 40m class telescopes. Second, increasing the number of modes and the aberration level increases the dimensionality of the problem. Thus, defining different datasets (with different dimensionalities), rather that a single one containing all the studied cases, allows to better understand the performance obtained, \ie fundamental limit for wavefront sensing versus limitations of the CNN models (e.g. generalization power or suboptimal training).
Nevertheless, in \sref{sec:dynamic}, we consider other appropriate datasets: one drawn from a uniform rms WFE distribution, and several with higher level of aberrations.

\subsection{Network architectures and training}

We consider two approaches to our problem, one where the CNN is trained to estimate Zernike coefficients, and one where the CNN is trained to do a direct phase map estimation.
During the onset of this work, we considered a number of architectures with good ranking at ImageNet classification challenges: VGG-16, Inception v3, ResNet-50, U-Net, and U-Net++. Eventually, and in this paper, we only use ResNet-50 and U-Net, the other architectures either did not work well for our application or do not add further insights to the topics discussed here. It is worth noting however that Inception v3 has shown promising results in different simulation studies \citep{Paine:18,Andersen:2019}.

Residual neural networks \citep{resnet}, or ResNet, are very deep networks where skip connections are introduced to improve gradient flow during the training steps. We use ResNet-50, which is 50 layers deep, and we initialize it with the parameters pre-trained on ImageNet. In order to adapt the architectures to the prediction of Zernike coefficients, the softmax activation and the last fully connected layers were replaced to match the output requirements.

For the direct phase estimation approach, we focussed on an architecture initially developed for biomedical image segmentation: U-Net \citep{unet}. The overall network structure follows a U-shaped geometry. The encoding part is made of successive 3$\times$3 convolution layers followed by 2$\times$2 max pooling layers. The input PSF images are thus progressively downsampled while the most relevant features are extracted. The contracting part is followed by an expansion part replacing pooling operators by upsampling operators. Importantly, there are skip connections combining features from the contracting path with the upsampling  part.
Since we perform regression rather than segmentation, the last softmax layer was removed.
In our implementation, the input PSF images and the output phase maps have the same grid sizes.

For the optimization we used Adam \citep{Kingma:14} with an initial learning rate of $10^{-3}$ and a scheduler dividing the learning rate by two every 75 epochs.
Our typical training procedure consists of a dataset of 100,000 entries, each consisting of two focal plane images and one phase map. Before being fed into the CNN, photon noise is added to the images, a square root stretch is applied, and each image is normalized by its maximum.
Our dataset is typically split in a 90-10 ratio, \ie 90,000 entries are used for training and 10,000 for validation.
We use a batch size of 64 entries, all batches constitute one epoch, and we train for 200 epochs.
The results and analyses, as presented in \sref{sec:results}, use different datasets based on different random seeds for the Zernike coefficient generation.
The loss function corresponds to the rms error, \ie
\begin{equation}\label{eq:loss}
\mathrm{loss} (\varphi, \hat{\varphi}) = \sqrt{ \dfrac{1}{N}  \sum_{i,j}^{N} \left(\varphi (x_i, y_j) - \hat{\varphi}(x_i, y_j)\right)^2},
\end{equation}
with $N$ the total number of pixels per phase map, $\hat{\varphi}$ the estimated phase and $\varphi$ the true phase map.

\section{Results and analysis}\label{sec:results}

In this section we explore the performance of our CNN models under idealized and degraded conditions.
Both architectures, ResNet-50 and U-Net, give similar results. Therefore we will only compare them when appropriate and we will use them interchangeably otherwise.

\subsection{Fundamental limit for wavefront sensing}\label{sec:limit}
The wavefront estimation is fundamentally limited by the information contained in the measurement process.
The Fisher information matrix and the Cramér-Rao (CR) bound are typically used to quantify the information ultimately extractable from the measurement whatever the estimation method. More specifically, the CR bound gives the lower bound on the error variance, and its reciprocal is the Fisher information for an unbiased estimator.
Several studies rely on this lower bound to get the fundamental limit of different wavefront sensor's performances \citep{Paterson:08,Paterson:13,Lee:99,Schulz:99,Noethe:07,Plantet:15}.

Considering photon noise only, \ie the ultimate noise limit since it pertains to the nature of light, \cite{Paterson:08} derives a fundamental limit for wavefront sensing without any assumption on the optics, if only that the wavefront sensor transforms the pupil phase in a measureable intensity. He derives the CR bound and finds that the measurement error of an aberration mode $j$ must satisfy $\sigma_j^2 \geq 1/(4 N_{\rm ph})$.
Interestingly, this limit can also be derived from the uncertainty principle in the form $\Delta \varphi ~ \Delta N_{\rm ph} \geq 1/2$, or $\Delta \varphi \geq 1 / \sqrt{4 N_{\rm ph}}$  for $N_{\rm ph}$ independent photon probes.
However, the existence of this bound does not guarantee that it is actually possible to reach it.
The most sensitive wavefront sensors, among existing concepts, are the Zernike wavefront sensor (ZWFS) \citep{N'Diaye:13,Guyon:05} and the iQuad \citep{Fauvarque:19}. Both concepts rely on a $\pi/2$ phase shift between different parts of the focal plane and differ in its tessellation.  Under photon noise only, the measurement error of the ZWFS\footnote{for a phase error close to zero and phase shift of $\pi/2$ over the central $1.06\lambda/D$ \citep{N'Diaye:13}.} is  $\sigma_j^2 = 1/ (2 N_{\rm ph})$  for the aberration mode $j$ \citep{N'Diaye:13}. 
 The loss of a factor two with respect to the fundamental limit is the result of the diffraction by the phase mask distributing half of the light outside the exit pupil. Since only the exit geometrical pupil is used for the measurement, the signal is effectively half of the input. Recently, however, a variation on the ZWFS with an enlarged central dot has been proposed improving its sensitivity at the expense of the lower spatial frequencies \citep{Chambouleyron:21}. In those conditions, the measurement error can almost reach the fundamental limit of $1/4 N_{ph}$.

Second after the ZWFS, focal plane wavefront sensing offers a high sensitivity with a measurement error known to be $\propto 1/N_{\rm ph}$   \citep[e.g.][]{Guyon:05,Meynadier:99,Bos:19,Paul:13}. Classical focal plane imaging does not maximize the phase contrast, like the ZWFS, which explains a loss of sensitivity compared to the fundamental limit \citep{Paterson:08}.
Nevertheless, FPWFS still provides a very high sensitivity across a wide range of spatial frequencies, in contrast to, e.g., the Shack-Hartmann wavefront sensor, which has a poor sensitivity at low spatial frequencies \citep[e.g.][]{Guyon:05}.
In practice, and for the analysis in this paper, the expected total theoretical residual error for $N_{\rm modes}$ statistically independent aberration modes is expressed by
\begin{equation}\label{eq:fundamental_limit}
  \sigma^2_{\rm th}  = N_{\rm modes}  \dfrac{1}{n_{\rm img} N_{\rm ph} }   \quad \textrm{[rad$^2$]},
\end{equation}
with $n_{\rm img}$ and $N_{\rm ph}$ the number of images and the number of photons per images respectively.

\subsection{Performance limit of CNNs}
To analyze the capability of our CNN in terms of sensitivity, we train and evaluate different models over a broad range of flux levels: from $10^2$ to $10^7$ photons per image.
This corresponds to a range of star magnitudes $m_H= 7 - 19.5$ (for $10^7$ and $10^2$ photons respectively) assuming the following parameters: an integration time of $T_i=1$s, a transmission and quantum efficiency equals to 50\%, a telescope diameter of 10m, and a filter bandwidth of 50nm.
We examine the two levels of aberrations (70nm and 350nm rms) and the two different spatial frequency contents (20 and 100 Zernike modes) described in \sref{sec:dataset}.
This allows us to study the limit of our trained models as a function of an increased dimensionality of the wavefront sensing problem.

\begin{figure}
  \includegraphics[width=\linewidth]{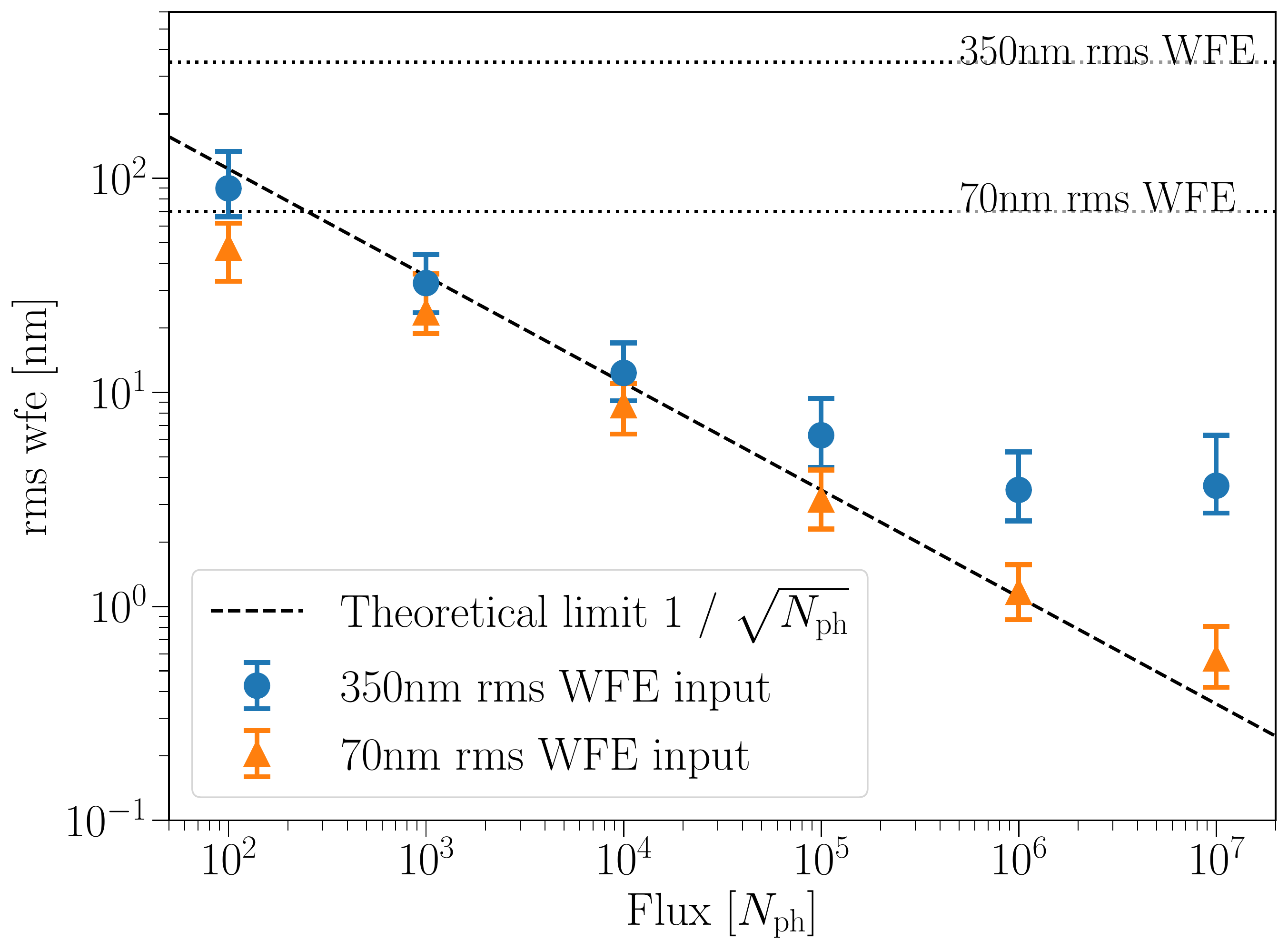}
  \includegraphics[width=\linewidth]{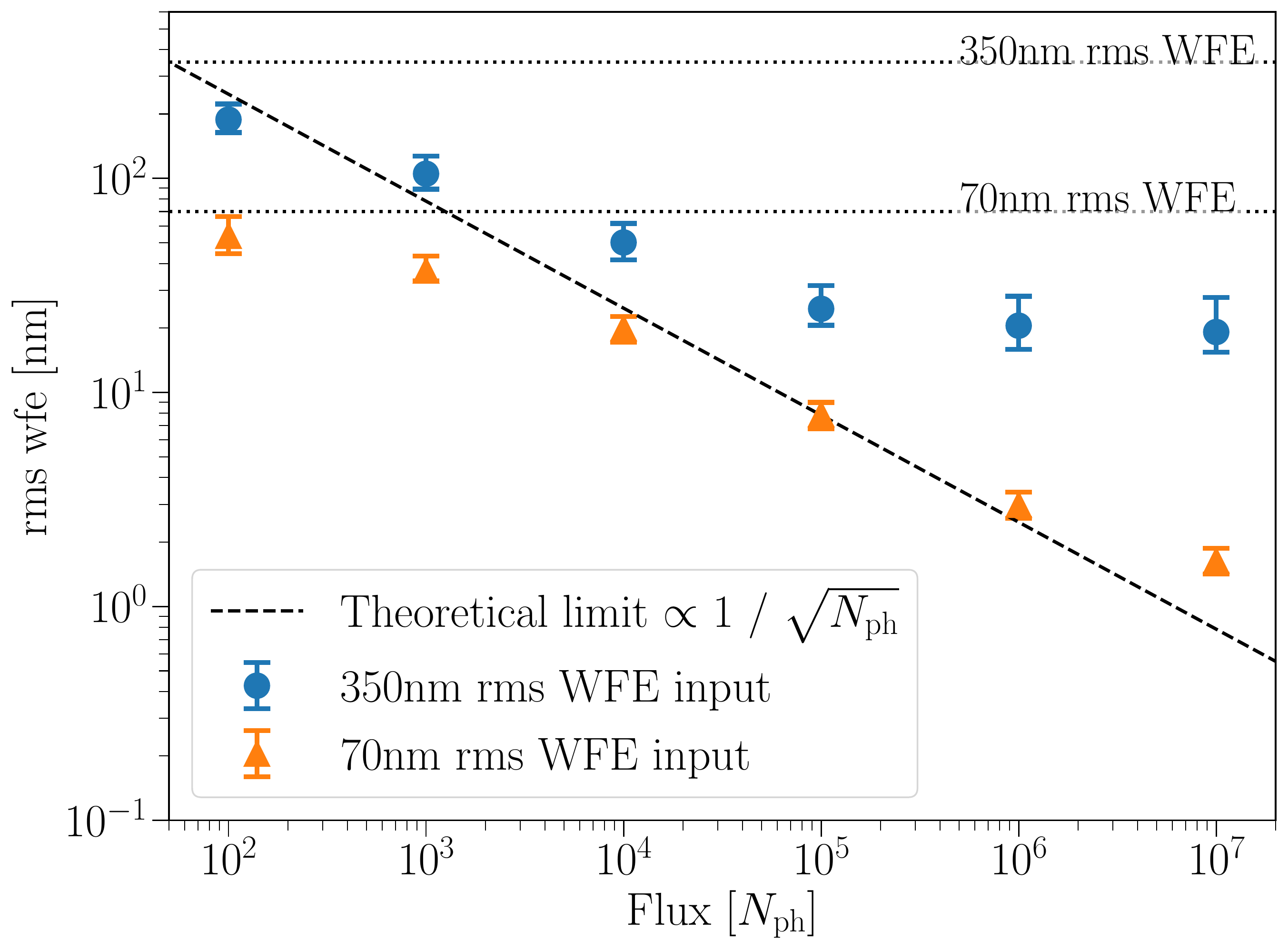}
  \caption{Residual rms WFE as a function of the flux per image, for an input median rms WFE of 70nm (orange) and 350nm (blue). Each point uses a model specifically trained for that flux and aberration regime. (Top) Low spatial frequency content with 20 Zernike modes. (Bottom) Higher spatial frequency content with 100 Zernike modes. }
  \label{fig:fundamental_limit}
\end{figure}

The results are illustrated in \fref{fig:fundamental_limit}.
Each point is the median residual error of 100 evaluations and the error bars are the 5-95\% percentiles\footnote{Note that this is applicable to all the following figures with error bars.}.
Note that the 100 evaluations refer to different phase screens, and not just to different photon noise realizations. These figures show where the performance is limited by photon noise and where it is limited by the model accuracy.

In the low aberration regime distributed over 20 Zernike coefficients, we can observe that the CNN reaches the sensitivity limit defined in \eref{eq:fundamental_limit} over a broad range of photon levels. The only exception is the low flux regime ($\lesssim1e4$) where the error does not become arbitrarily large but reaches a saturation level at around 70nm rms.
 A similar saturation level is observed at low flux with COFFEE, a coronagraphic phase diversity method based on a maximum a posteriori approach, when the appropriate regularization is used \citep{Paul:13}.
 In our case, this saturation can be interpreted as an implicit regularization originating from the training data distribution. The limit is reached when no aberration can be distinguished from the noise, in which case the predicted phase tends to zero.
When the level of aberration is increased to 350nm WFE, we can observe a plateau at the high signal end.  This saturation level is a numerical limitation and can be reduced by increasing the dataset size. See also \sref{sec:size} for a dedicated discussion.

The same analysis is performed for 100 Zernike modes, see \fref{fig:fundamental_limit} (bottom). The low aberration regime exhibits a very similar behavior, while the high aberration regime case  is  more strongly influenced  by the saturation level due to a suboptimal training. Hence, the accuracy rarely reaches the theoretical one. Again, this can be mitigated by increasing the dataset size.

\subsection{Robustness to changing signal-to-noise ratio}\label{sec:noise}
\begin{figure}
  \centering
  \includegraphics[width=\linewidth]{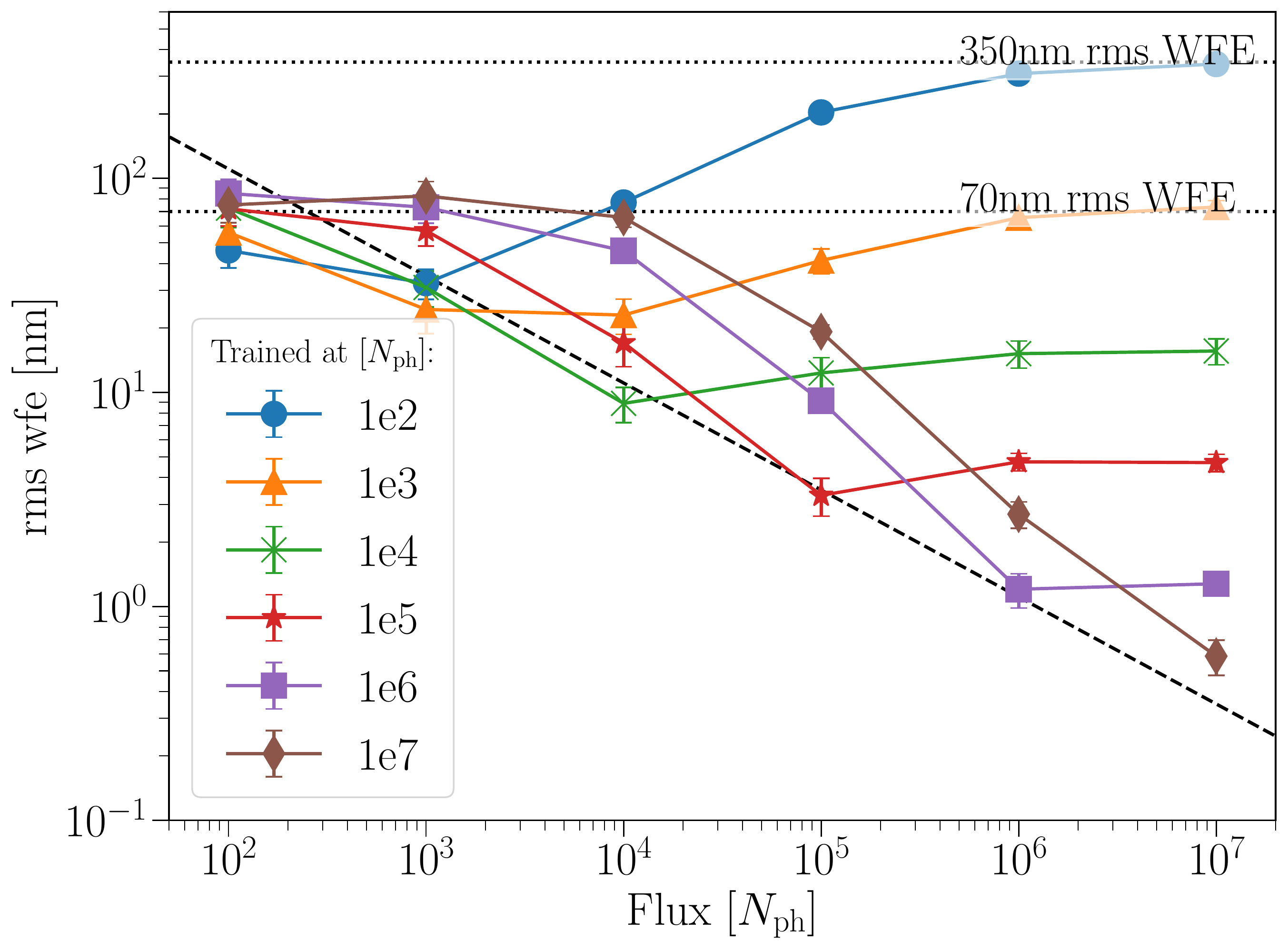}
  \caption{Robustness to changing photon flux levels. Each curve uses a different model  and is evaluated at six different flux levels. We use here ResNet-50 trained on datasets with input median rms WFE of 70nm rms distributed over 20 Zernike modes.}
  \label{fig:snr}
\end{figure}

In this section we explore the robustness of the CNN models under varying signal levels.
The network architectures are trained at a specific SNR and then, during evaluation, exposed to a range of SNR.
\Fref{fig:snr} is the result of six different trainings and 36 different evaluations, for a median WFE of 70nm rms and 20 Zernike modes.
At low signal level, the performance moves progressively away from the photon noise limit as the SNR used in the evaluation decreases, and converges to 70nm rms due to the intrinsic noise regularization.
At high signal level, the performance first degrades slightly as the SNR used in the evaluation increases, and then stagnates to a given WFE level, which depends on the training SNR. In all cases, the minimum WFE is reached at the training SNR.
It is noteworthy that the performance degradation is  mild in the vicinity of the training signal level, so that a single model might suffice for a range of observing conditions.

If robustness under a wide range of SNR levels is desired, the training could be adapted. The distribution of flux in the training dataset should be established based on the range of expected stellar magnitudes versus desired accuracy, and observing variability.
Alternatively a number of models could be used and selected as a function of the current stellar magnitude.

\subsection{Dynamic range}\label{sec:dynamic}
\begin{figure}
  \centering
  \includegraphics[width=\linewidth]{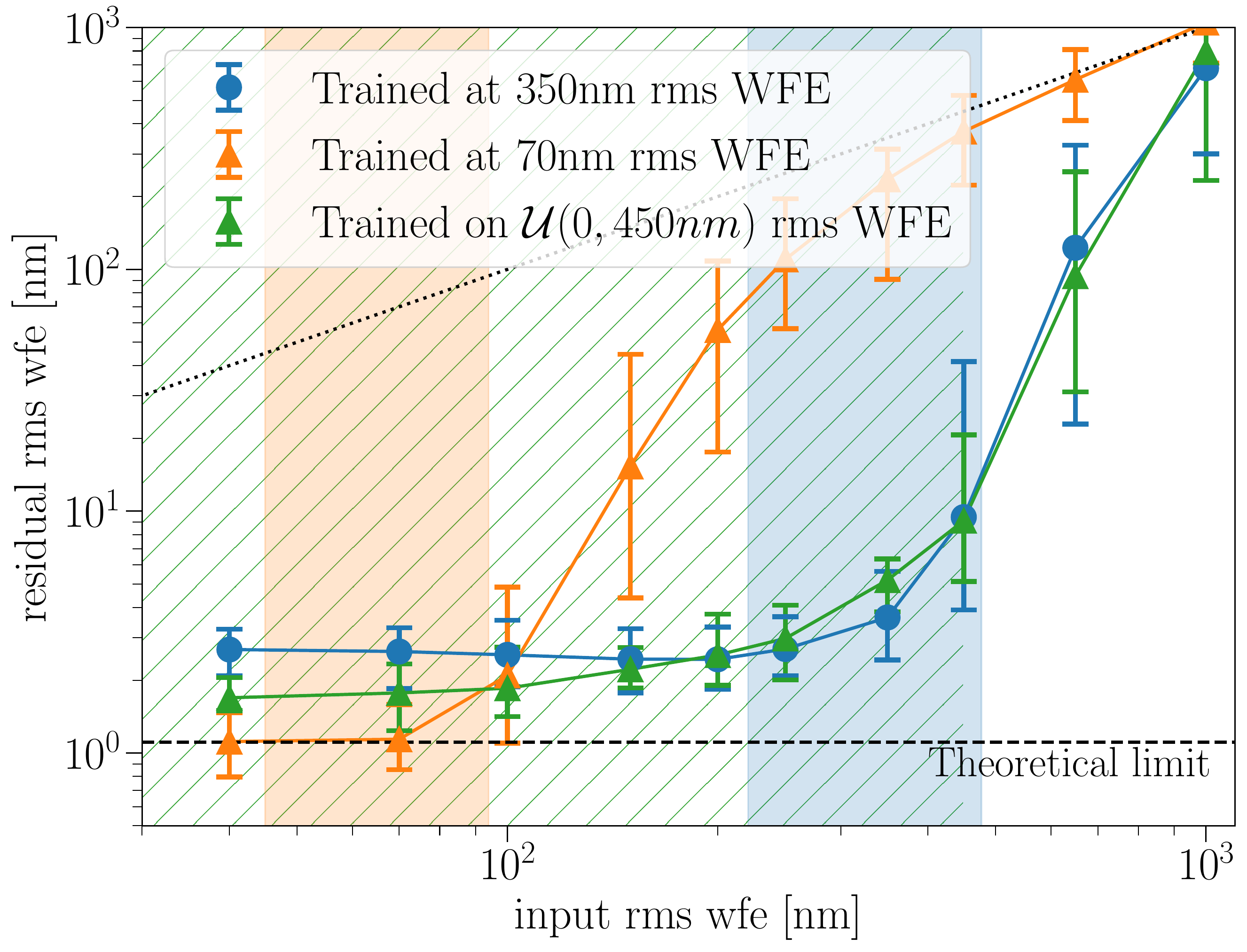}
  \caption{Residual rms WFE as a function of input rms WFE (40 to 450nm rms WFE) distributed over 20 Zernike modes. In blue, for a ResNet-50 model trained around 350nm WFE, in orange around 70nm WFE, and in green trained on a uniform distribution of wavefront aberrations ranging from 0 to 450nm WFE. The dashed line gives the fundamental limit as discussed in \sref{sec:limit}. The dotted line gives the one-to-one relation. The shaded areas represent the 2-98\% rms WFE percentiles in the respective training datasets.}
  \label{fig:dynamical}
\end{figure}

Here we analyze the accuracy of the estimated wavefront for different levels of aberrations. In \Fref{fig:dynamical} we use two different models trained at low  (70nm rms) and high  (350nm) aberration levels. We test those models on eight datasets with aberration level ranging from 40nm up to 450nm rms, corresponding to  0.11 up to 1.28 rad rms. The results show a similar or better accuracy for aberration level below the trained one, and a rapidly decreasing accuracy at higher aberration levels.

The behavior at low aberration level results from the way our datasets are generated, where each Zernike coefficient is drawn from a uniform distribution around zero. Hence the training implicitely includes low-aberration samples.
It is interesting to note that the evaluation at the trained aberration level (70nm and 350nm resp.) is actually not where the best performance can be found. Again this may be explained by the way our datasets are generated, with a distribution of WFE around a given median. In the case of a median WFE of 350nm rms, the aberration distribution extends down to $\sim$225nm (see blue shaded area in \Fref{fig:dynamical}), which approximately coincides with the lowest residual error in \Fref{fig:dynamical}.
The performance is here limited by the size of the dataset (100,000 entries), and since the effective number of entries with $\geq 350$nm WFE is simply lower than for $\geq 225$nm WFE in the training dataset, the results are better at lower aberrations than at the median WFE.

We then train a model on a uniform distribution of wavefront aberrations ranging from 0 to 450nm rms. To obtain a good training, we set the weight decay of the Adam optimizer to 1e-7, which is otherwise set to 0 by default.
The result is also shown in \Fref{fig:dynamical}. We observe an intermediate behavior where the model performs better at low aberration but slightly worse at higher aberration compared to the model trained at 350nm rms.

At higher aberration level, the models are less efficient in picking up features and extracting useful information.
Nevertheless, it is worth noting that a valid correction extends well beyond the trained aberration level, and despite a rapidly decreasing accuracy.

\begin{figure}
\begin{minipage}{\linewidth}
  \begin{minipage}[t]{1\linewidth}\vspace{0pt}
  \includegraphics[width=\linewidth]{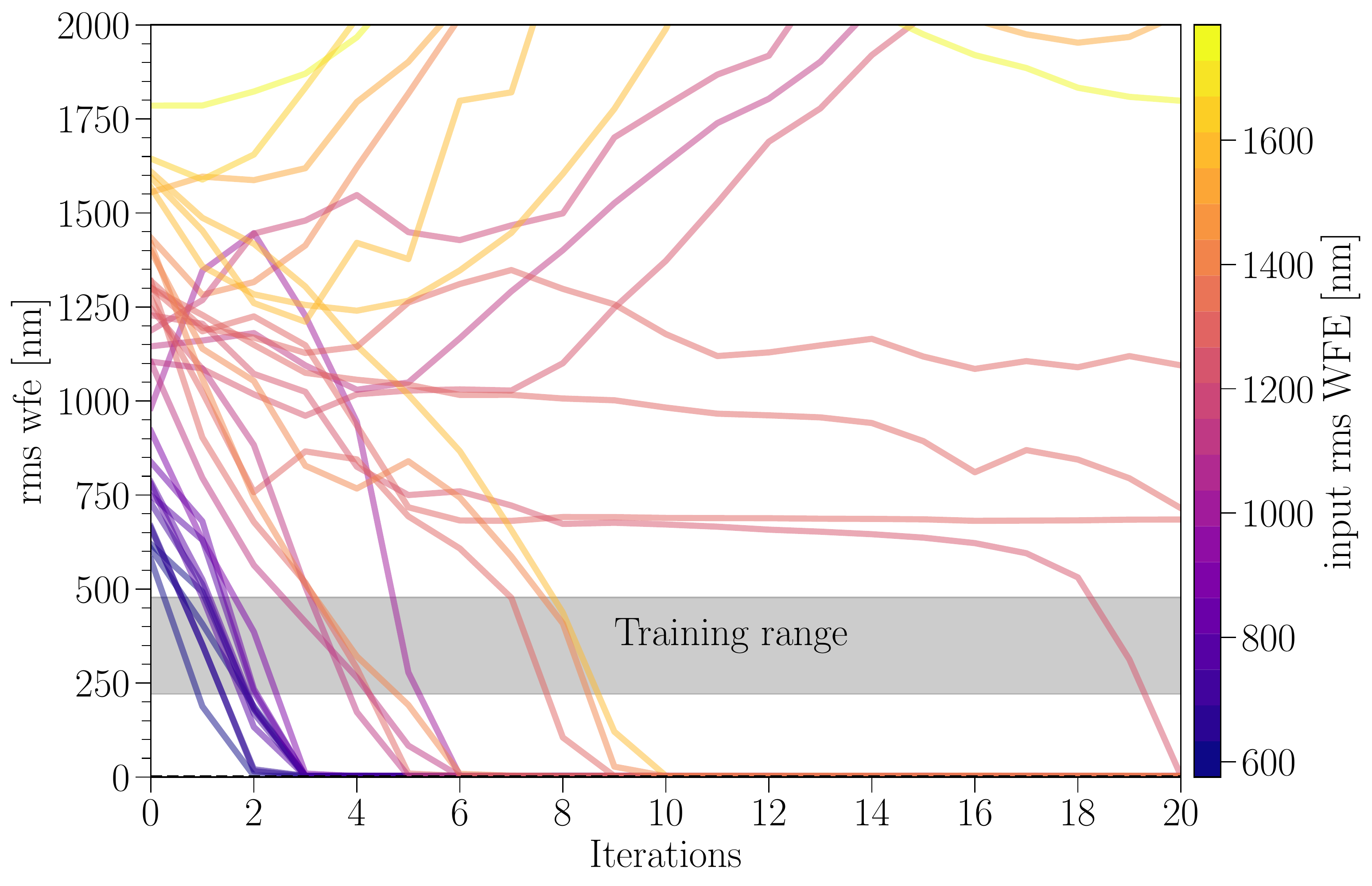}
  \end{minipage}
  \begin{minipage}[t]{1.1\linewidth}\vspace{2.5pt}\raggedright
    \hspace{0.cm}
  \includegraphics[width=0.45\linewidth]{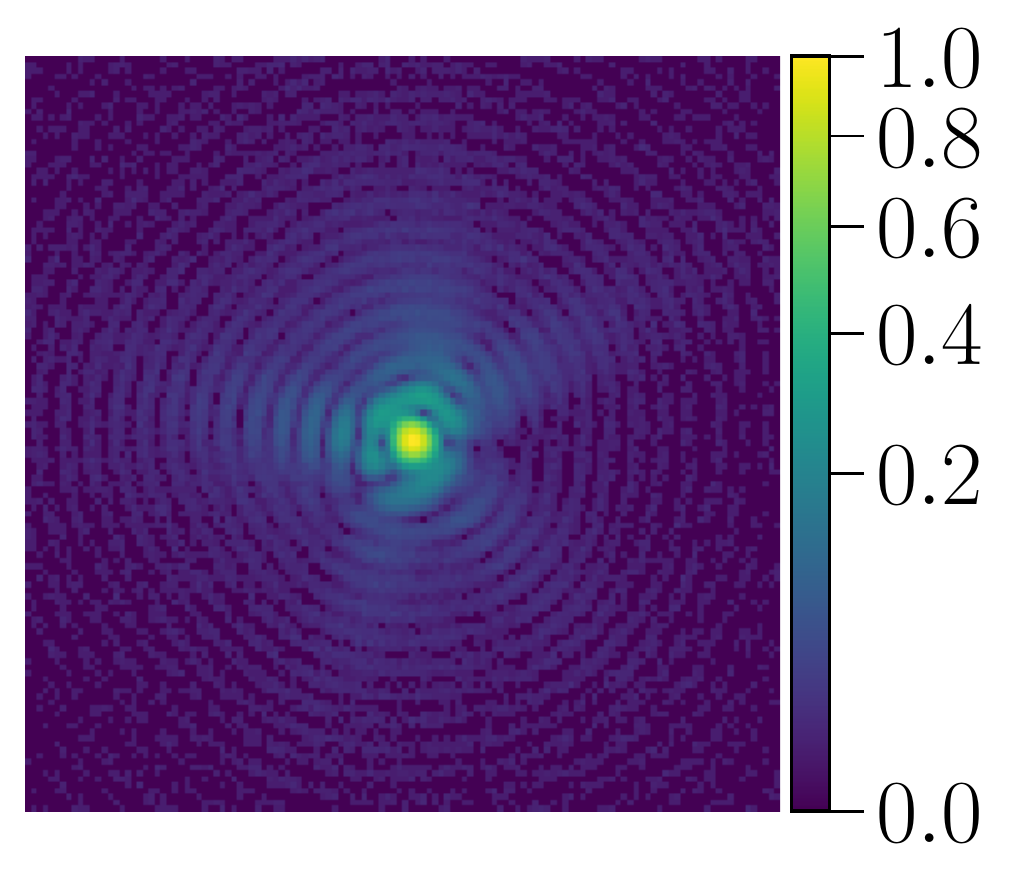}
  \includegraphics[width=0.45\linewidth]{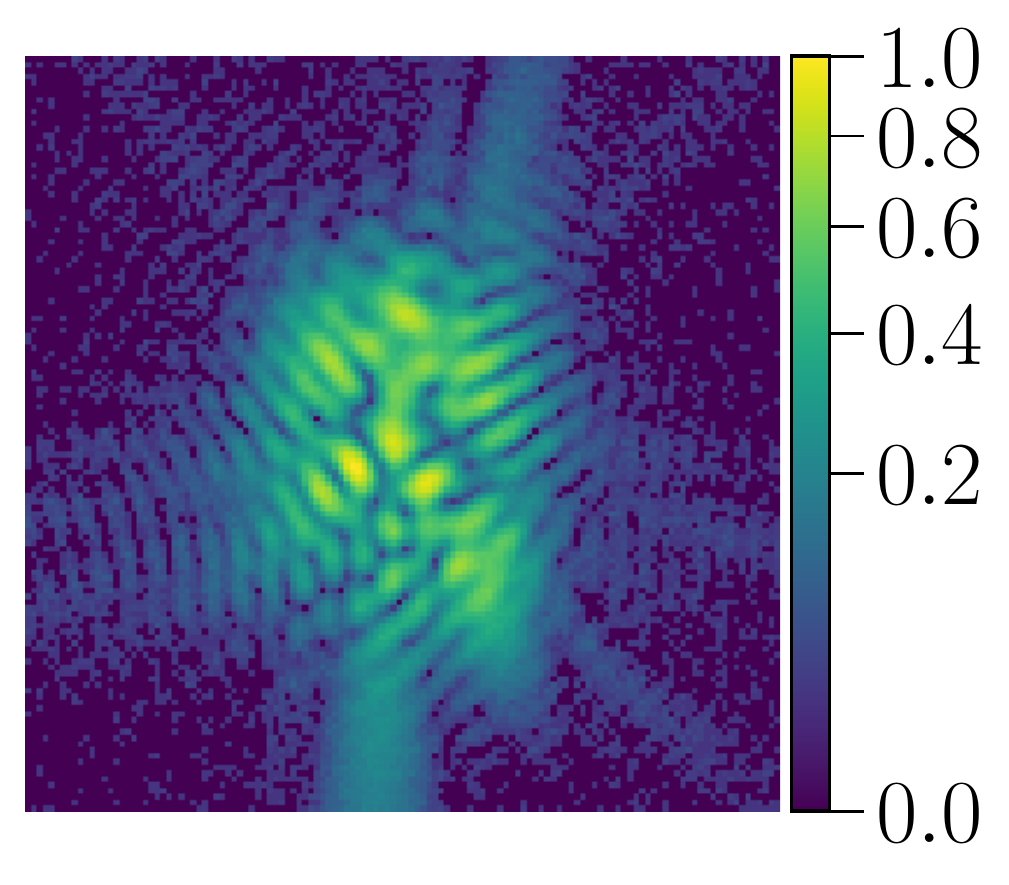}
  \end{minipage}
\end{minipage}
  \caption{(Top) Iterative application of the CNN to different level of aberrations. (Bottom) Illustration of two PSFs, with rms WFE of 320nm (top) and 1060nm (bottom), in both cases the CNN converges after a few iterations.}
  \label{fig:CNN_iterative}
\end{figure}

Following the results in \fref{fig:dynamical}, we apply iteratively the CNN model for aberrations well beyond the trained aberration level. At each iteration, the CNN infers a wavefront, which we subtract from the input to produce a new pair of PSFs based on the residual wavefront.
We test this iterative approach for 40 different aberration levels ranging from 500nm to 1750nm.
To limit the effect of field-of-view cropping, the tip-tilt modes are removed from all phase maps (and the quoted WFE are also calculated without tip-tilt). The training range of the CNN is here around 350nm rms.
The results are presented in \fref{fig:CNN_iterative}.
We can observe that the CNN properly converges in a few iterations for initial aberration levels well beyond its training range. Also, once it has converged, the correction stays stable. It is only for initial aberration levels $\gtrsim1.1\mu m$ that the rms WFE either stagnates or starts to diverge.
In \fref{fig:CNN_iterative} (right) two PSFs are illustrated, with  rms WFE of $\sim 320$nm and $\sim 1060$nm respectively. Both are well corrected after a few iterations.  The morphology of those images is very different, yet the CNN is able to converge, which is remarkable.
Overall these are very encouraging results for real applications, typically running in closed-loop, starting with high levels of aberrations and with the objective of stabilizing the wavefront error to   the lowest level.

An alternative approach would be of course to train the CNN model on a wider aberration range for which the application would not need to be iterative. As a comparison, we trained three additional CNNs\footnote{with 100,000 entries in our dataset, and for a photon flux of 1e6.} with larger aberrations~: 535, 800, 1070nm rms (or 700, 1050, and 1400 if the tip-tilt was not removed), and we test them on a range of aberrations from 250 to 1070nm, similarly to the results presented in \fref{fig:dynamical}. The bottom line of this comparison is that the residual wavefront error increases with the level of aberrations the CNN was trained with, in particular for training at 350, 535, 800, 1070, the residual error is 3.5, 5.2, 13.6, 27.4nm rms. While the training can certainly be improved with, e.g., larger datasets or a better distribution of the input WFE in the datasets, the results presented in \fref{fig:CNN_iterative} show that we can also benefit from the generalization power of CNN, for instance, for the bootstrapping phase in a closed-loop system, without the need for more demanding training strategy or different network architectures, such as recurrent neural networks.

\subsection{Robustness to higher-order disturbances}\label{sec:high order}

\begin{figure*}
  \includegraphics[width=\columnwidth]{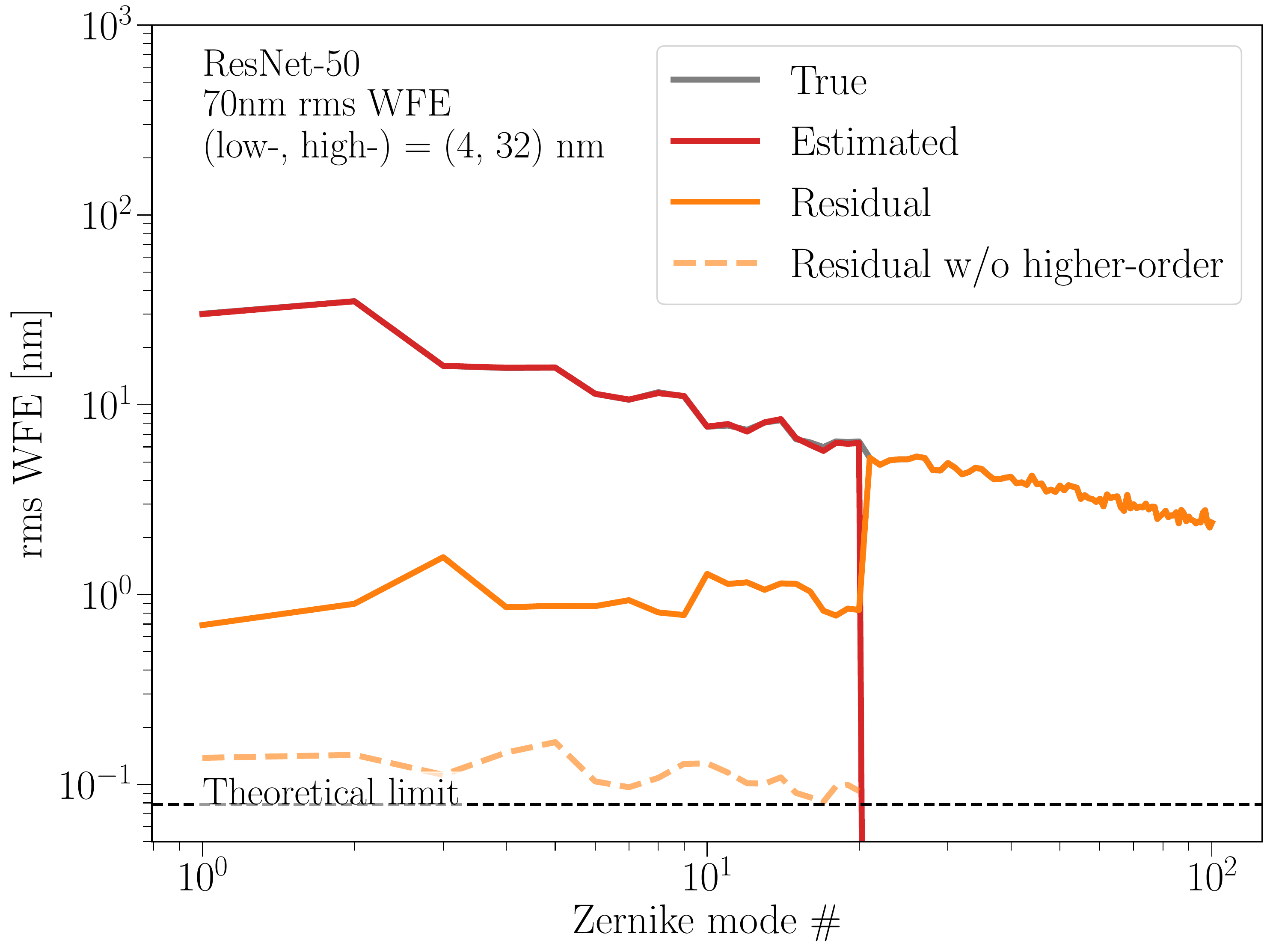}
  \includegraphics[width=\columnwidth]{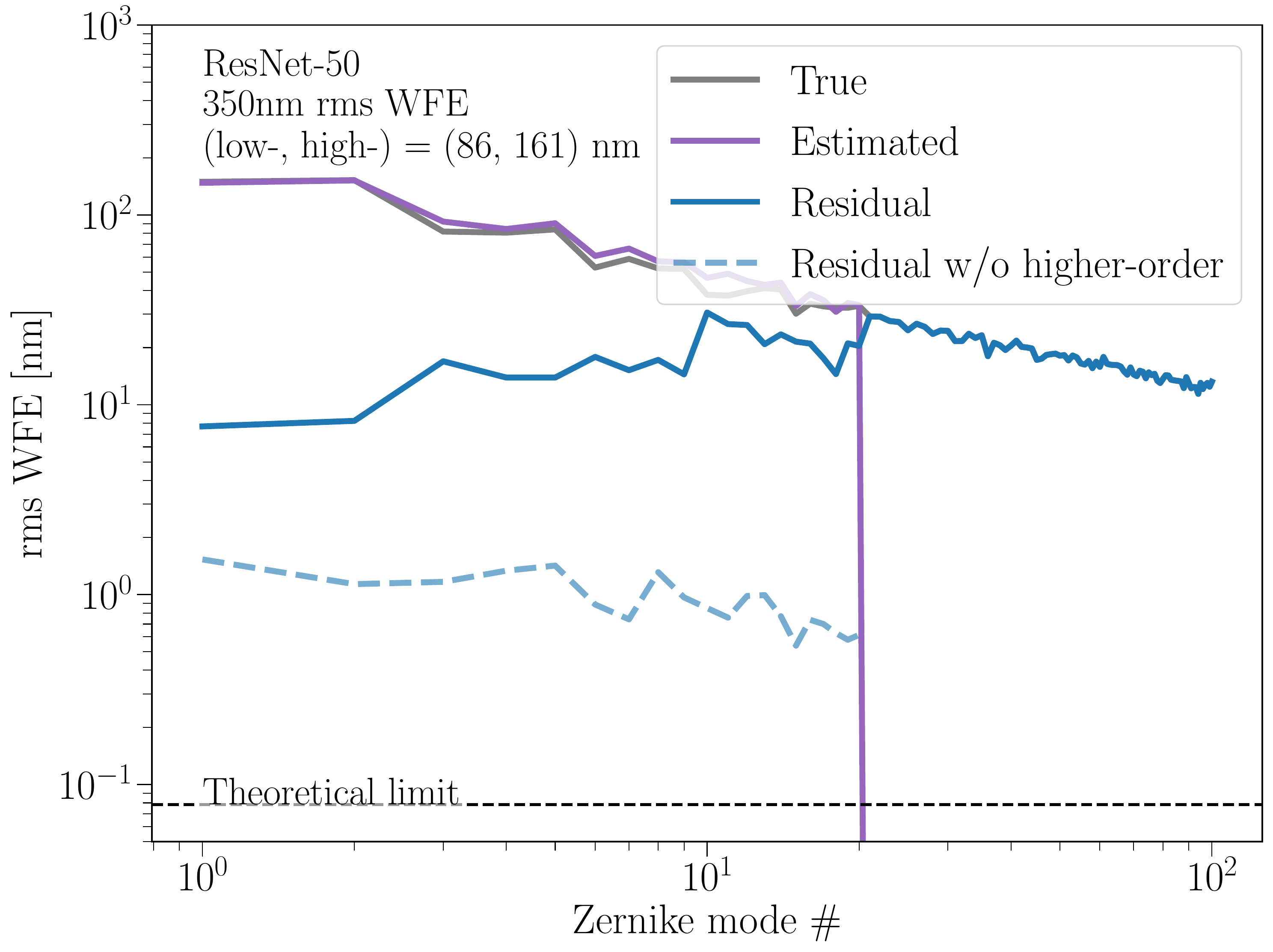}
  \includegraphics[width=\columnwidth]{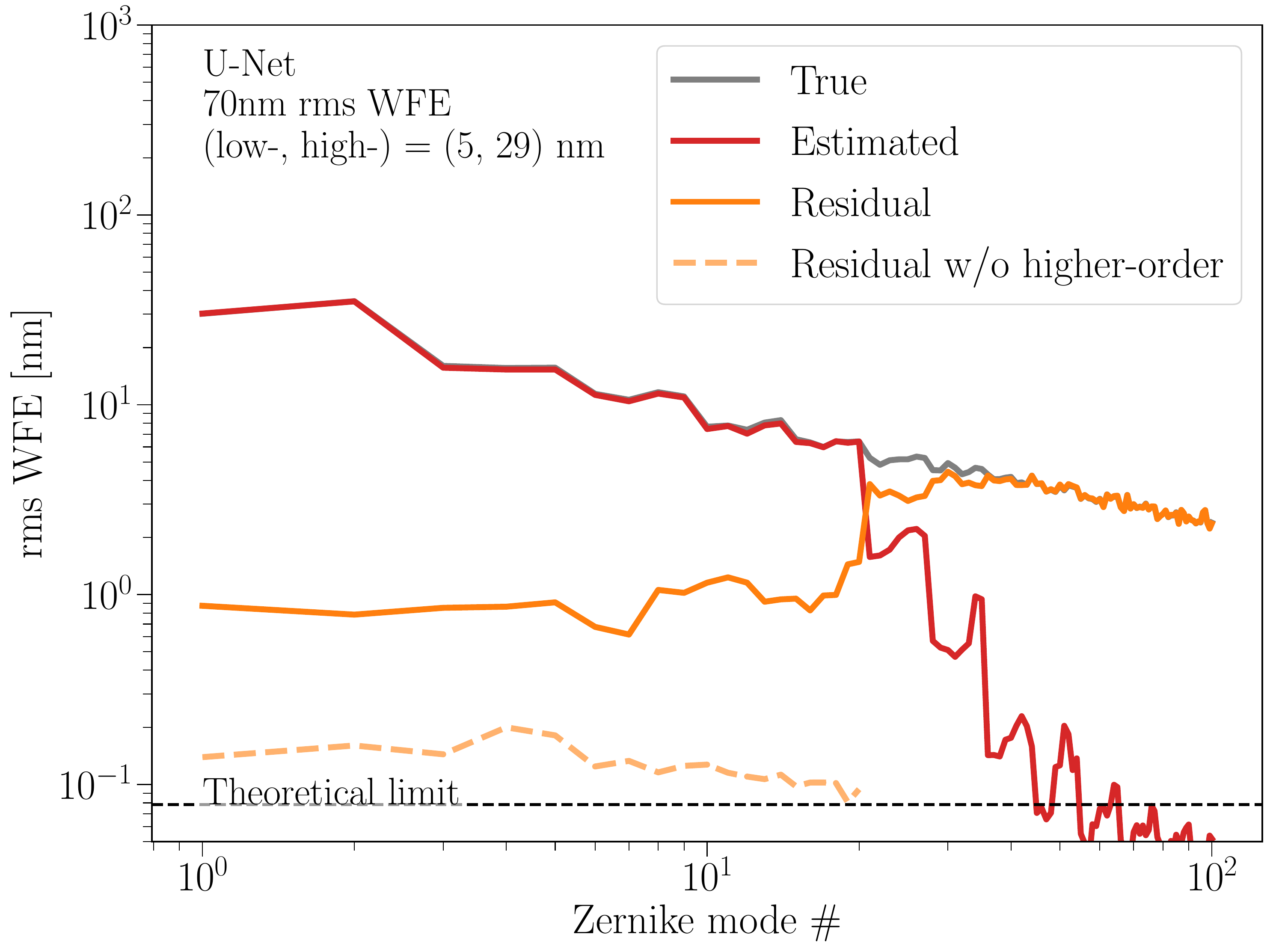}
  \includegraphics[width=\columnwidth]{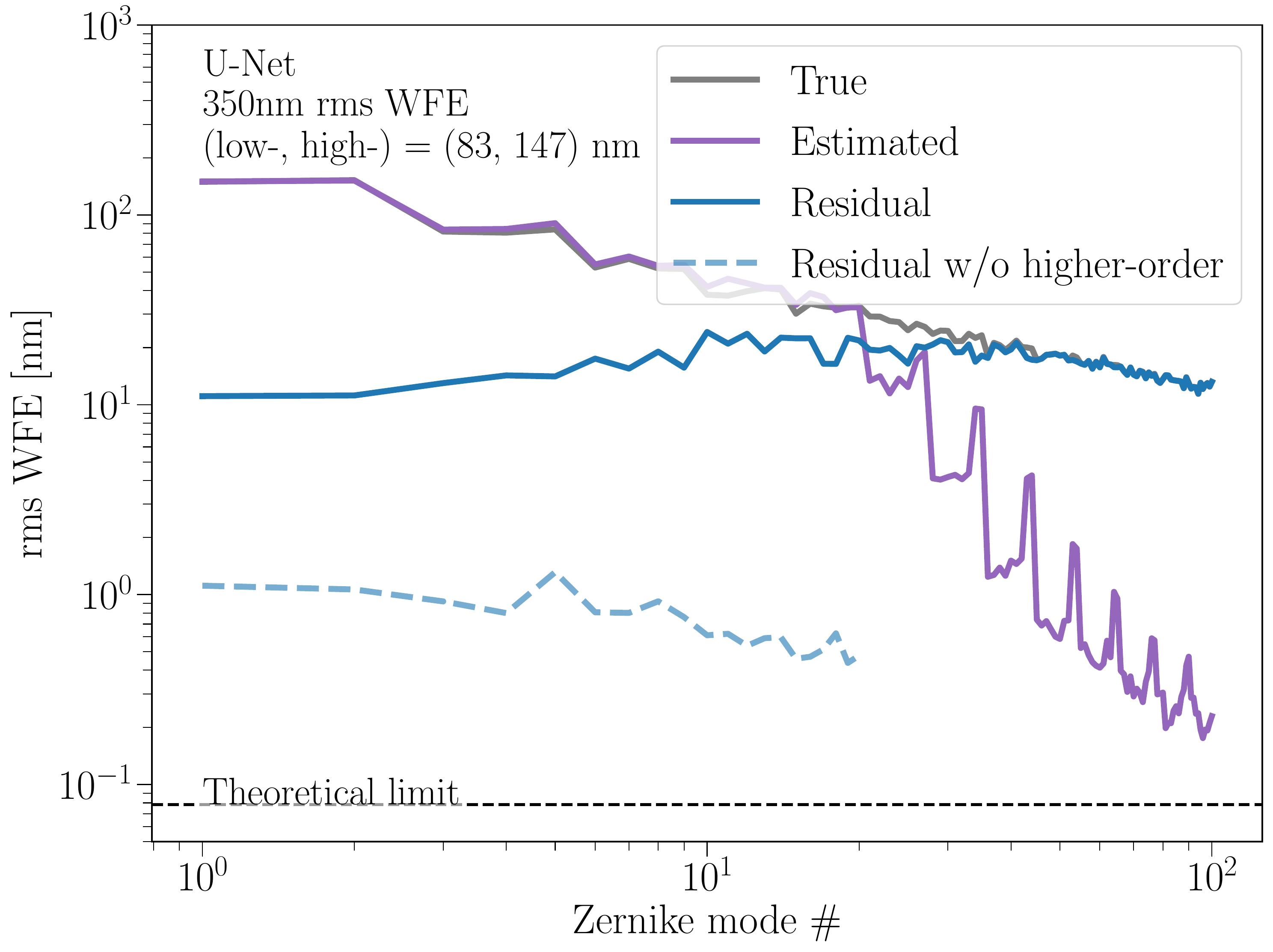}
  \caption{Modal rms wavefront error  for two different levels of aberrations: 70nm rms (Left; red and orange) and 350nm rms (Right; purple and blue); and two architectures: ResNet-50 (Top), U-Net (Bottom). The models are exposed to aberrated PSFs
with higher-order aberrations compared to the training dataset (solid lines). The estimated phase rms WFE (red and purple) and the residual errors (orange and blue) are compared to the true phase rms WFE (grey).
The residual errors, when no higher-order aberrations are present, are also plotted in dashed lines.
   In all cases the signal per image is $10^7$ photons. For a median input of 70nm rms, both ResNet and U-Net reduce the WFE of the low-order modes to about 4nm rms, while the higher-order modes are essentially unaffected and have an rms of about 30nm. When the aberrations are larger (350nm rms), the models start to be significantly affected and the rms error increases to about 80nm (140 nm) for the low- (high-) orders respectively. }
  \label{fig:ho}
\end{figure*}

In realistic conditions, a wavefront sensor measures the projection of the phase aberration on a finite set of values, such as Zernike coefficients or zonal values corresponding to deformable mirror actuators. Higher-order aberrations are thus not sensed and can be considered as an additional source of disturbance. This is the case, for example, with the Shack-Hartmann wavefront sensor, where a limited sampling of the wavefront leads to an aliasing effect  increasing the measurement error.
It is therefore of practical interest to study how this unwanted signal affects the inference of the trained CNN.

To explore the effect of higher-order disturbances, we use different models trained on phase maps constructed with 20 Zernike modes and evaluate their accuracy on a dataset with 100 Zernike modes.
We consider both ResNet-50 and U-Net models, with input WFE of 70 and 350nm rms.
The residual wavefront errors ($\Delta \varphi =  \varphi - \hat{\varphi}$) are projected on 100 Zernike modes and the rms (over 100 different evaluations) of the obtained Zernike coefficients are calculated.
The results are illustrated in \Fref{fig:ho}.
We can observe a higher loss of accuracy for higher aberration levels, see \Fref{fig:ho} (left) versus (right), and for higher flux levels (not shown). The simple interpretation is that the CNN models are more affected by higher-order aberrations as they become more prominent, and therefore distinctive, with respect to the photon noise.
Although a minor effect, it is interesting to note that the direct phase estimation done by U-Net provides a valid correction beyond the 20 Zernike modes for which it was trained, while the ResNet-50 is bounded to the first 20 Zernike modes by construction.

While for NCPA correction, behind an extreme adaptive optics system where residual atmospheric aberrations are kept to a minimum, our training strategy might be appropriate, for adaptive optics application this degradation might be a showstopper (although we do not fully explore this here).
In fact, beyond the mild robustness offered by the CNN-based models, these results illustrate the following rule-of-thumb:  the training data should always be as representative as possible of the real observing conditions.

\subsection{Phase diversity: implication of sign ambiguity}\label{sec:diversity}

To properly recover, and avoid any ambiguity on the phase in the pupil plane using focal plane images requires a unique intensity measurement for a given phase aberration. This uniqueness is not generally guaranteed, in particular for circularly symmetric pupils.
Ambiguities occuring in phase retrieval is extensively discussed in the literature for applications from image reconstruction to wavefront sensing \citep[see e.g.][for a review relevant to astronomy]{Bos:19}.
The non-uniqueness in the case of circularly symmetric pupils is the well-known sign ambiguity, which results from the Hermitian properties of  Fraunhofer propagation.
Indeed, the pupil-plane electric field $E_{\rm pup} (x)$ and the same field flipped and conjugated $E_{\rm pup}^* (-x)$ have the same Fourier transform, and therefore lead to the same intensity distribution in the focal plane.
For an even phase aberration (defined by $\varphi_{\rm even}(x) = \varphi_{\rm even}(-x)$), and assuming an even amplitude distribution across the pupil -- which we omit in the following -- we can write
\begin{align}
E_{\rm pup}^* (-x) &= \exp(-j \varphi_{\rm even} (-x))\\
             &= \exp(-j \varphi_{\rm even}(x) ),
\end{align}
and thus
\begin{equation}
    \mathcal{F}\{\exp(-j \varphi_{\rm even}(x))\} = \mathcal{F}\{\exp(j \varphi_{\rm even}(x))\},
\end{equation}
\ie $\varphi_{\rm even}(x)$ and $- \varphi_{\rm even}(x)$ produce the same PSFs.
Expressing $\varphi_{\rm even}$  as a sum of even Zernike modes,  $\varphi_{\rm even} = \sum_{n,m; n =\text{ even}}  a_{n,m} Z_{n,m}$, one easily understands that the relative sign between even modes is not ambiguous and the degeneracy reduces to a single sign ambiguity, regardless of the number of even modes to be evaluated.

\begin{figure}
  \centering
  \includegraphics[width=\linewidth]{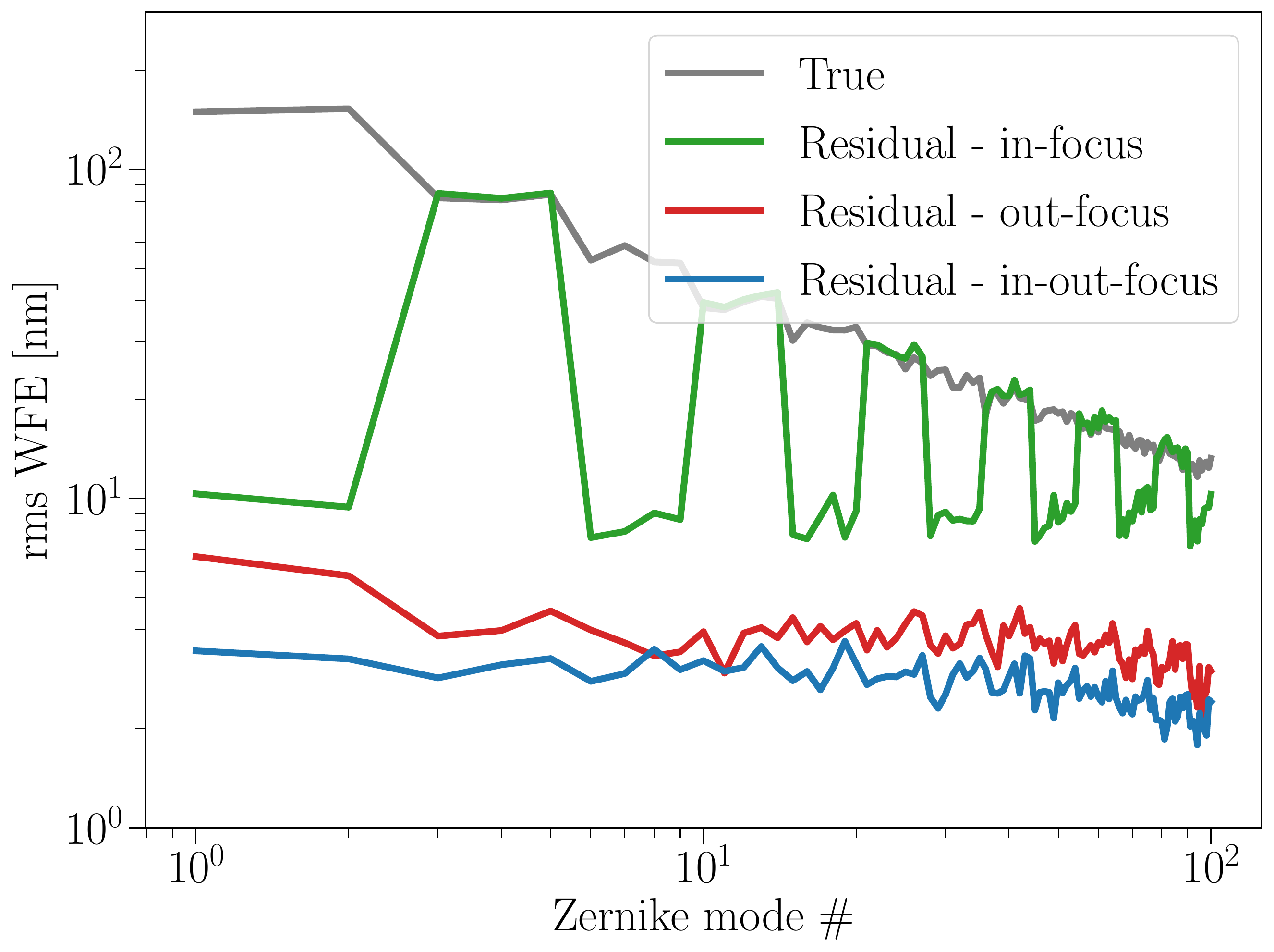}
  \caption{Modal wavefront error evaluated over 100 different phase maps. In blue, the input phase map following a spatial power spectrum with a slope of $-2$. The other curves give the residual error for three different models: using a single in-focus PSF, using a single out-of-focus PSF, and using both PSFs.}
  \label{fig:ambiguity}
\end{figure}

Many approaches exist to lift this ambiguity \citep[see e.g.][for high contrast imaging]{Jovanovic:18}.
Two natural (but not necessarily desired in the way they may affect the observations) ways are either to introduce a phase diversity as done in this paper, or to use a non-centro symmetric pupil support \citep[e.g.][]{Bos:19}.
Here we simply illustrate how the CNN behaves with respect to this sign ambiguity, by comparing the modal rms error for CNN models trained using two images (as in the previous sections), one defocussed image only, and one in-focus image only. The results are shown in \fref{fig:ambiguity}.
In the case of a single in-focus PSF, we see that the residual errors on the even modes correspond to the input error (the estimation is close to zero) as a result of the ambiguity, while odd modes are properly sensed.  However, the accuracy in the odd modes is degraded compared to the single defocussed case. This is the result of a suboptimal training and translates in  overfitting, which can be identified by the training and validation curves. This could be circumvented by adapting the loss function and replacing  \eref{eq:loss} by the rms error on the odd coefficients only.
The single defocussed PSF case does not suffer from sign-ambiguity. This is the result of an implicit prior: the introduced defocus is equal to $\lambda / 4$ or 550nm rms, while the focus term in the aberration to be sensed is drawn from a uniform distribution in the range $\sim[-150, 150]$nm rms (for 350nm rms WFE and 20 Zernike modes), so that the total focus term is always positive. The factor $\sqrt{2}$ between the single defocussed image and the two images case is solely due to the increased SNR.

\subsection{Effect of field-of-view and PSF sampling}\label{sec:pxscale}
In the results presented in the previous sections, the PSF was sampled with 4.5 pixels over $\lambda / D$, or a pixel scale of 0.22$\lambda/D$/px, with a grid size of 128$\times$128 pixels.
Here we study how the PSF sampling influences the performance. In order to preserve the exact same network architectures, we keep a fixed grid size of 128$\times$128 pixels. Therefore increasing the PSF sampling means reducing the field-of-view and potentially leaving out information. Conversely, a coarser sampling, in particular below the Nyquist sampling (\ie $<$2 pixel/$\lambda/D$), may lead to a loss of information.

\begin{figure}
  \includegraphics[width=\linewidth]{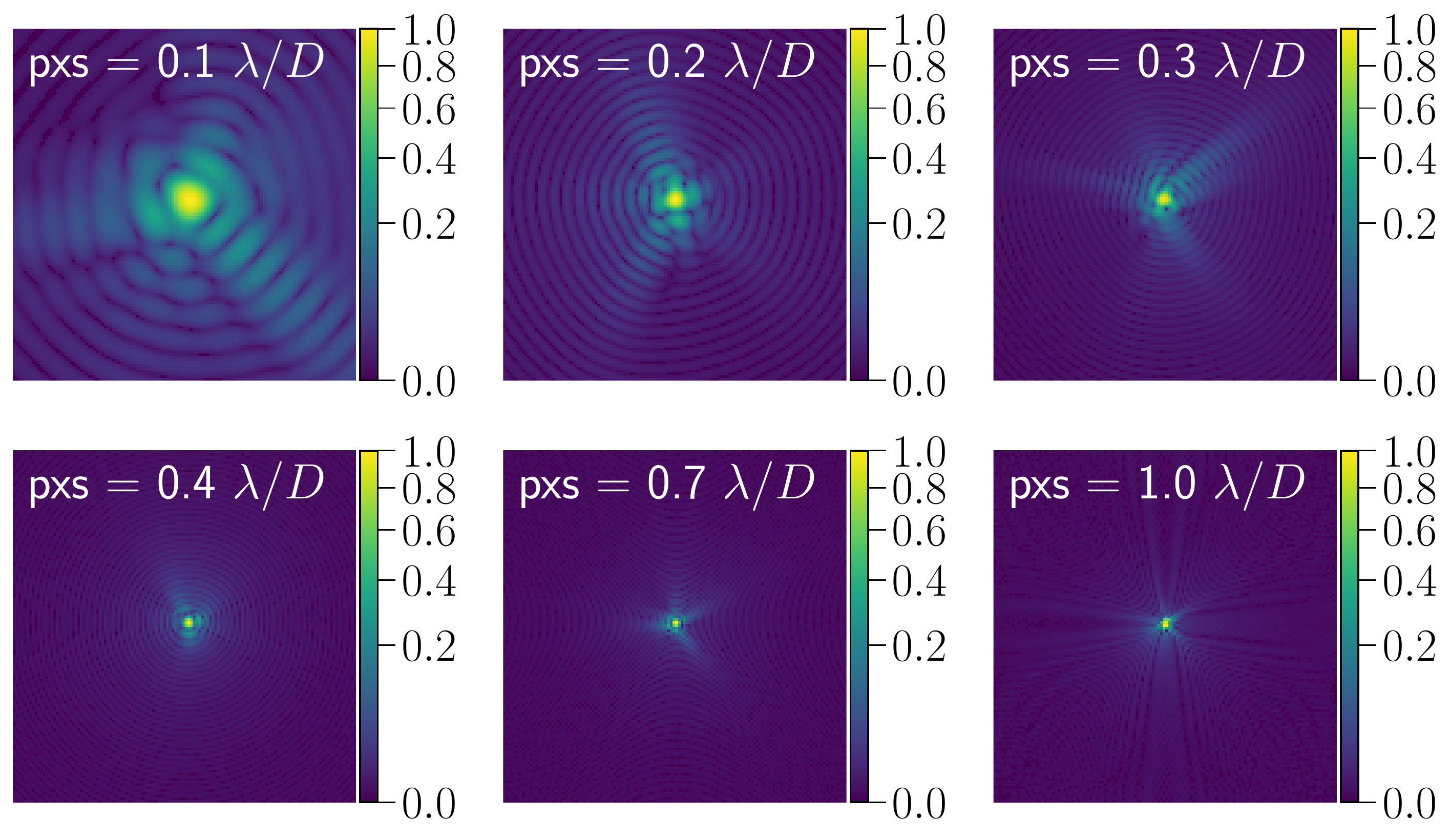}
  \caption{Illustration of different PSF sampling. Starting from (Top Left), the pixel scale is [0.1, 0.2, 0.3, 0.4, 0.7, 1] $\lambda/D$. Since the PSF gridsize is kept fixed to 128$\times$128 pixels, the field-of-view changes accordingly with, respectively, [12.8, 25.6, 38.4, 51.2, 89.6, 128] $\lambda/D$. }
  \label{fig:psfs_pxscale}
\end{figure}

\begin{figure}
  \centering
  \includegraphics[width=\linewidth]{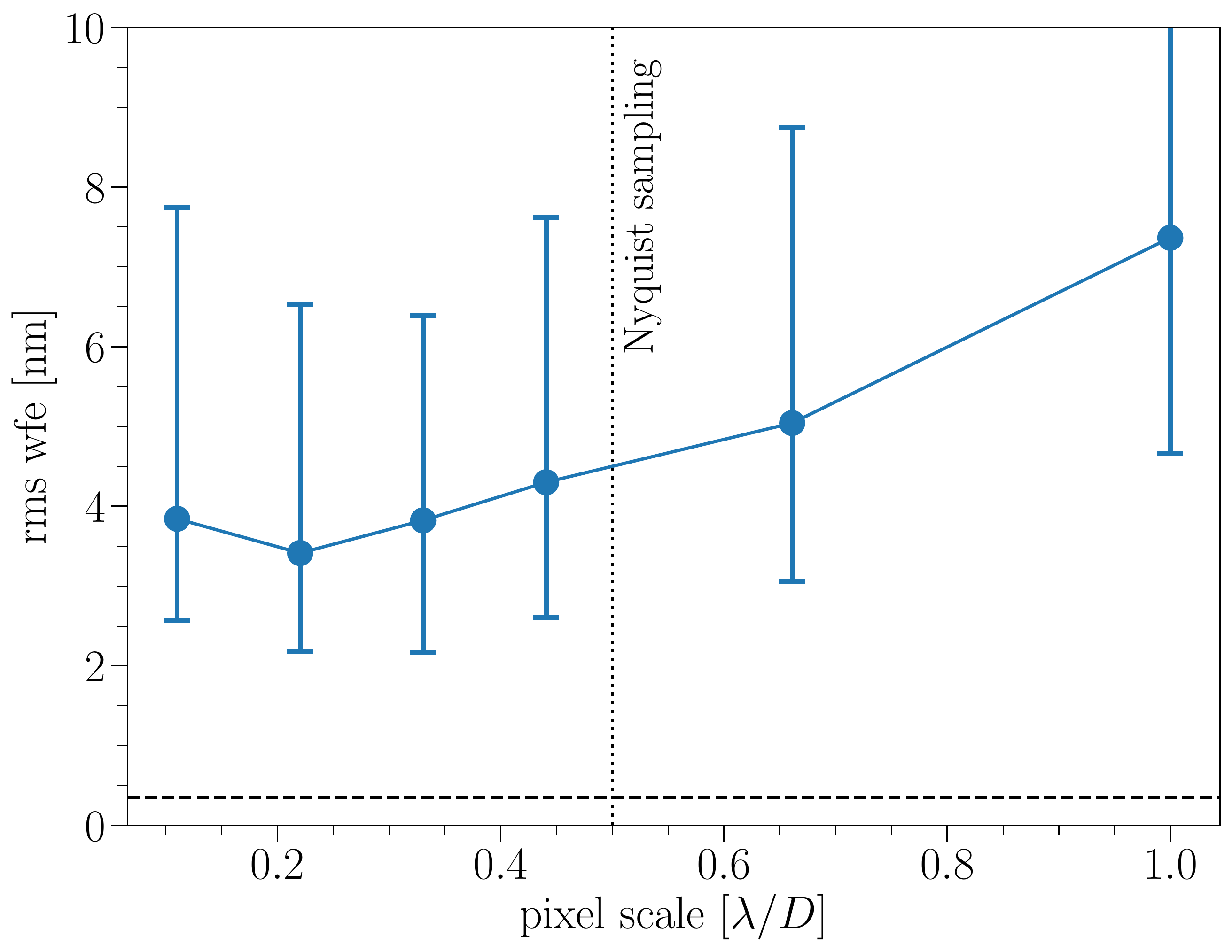}
  \caption{Residual rms WFE for different pixel scales. The same pixel scale is used for training and evaluation. Each point is a different model. The Nyquist sampling (0.5$\lambda/D$) is indicated by the vertical dotted line. The photon noise limit is indicated by the horizontal dashed line.}
  \label{fig:pixel_scale}
\end{figure}

To examine this effect, we generate different datasets with pixel scales between 0.1$\lambda/D$ and 1$\lambda/D$ and we train a new model for each case. An example of the generated PSFs is given \Fref{fig:psfs_pxscale}.
The results are illustrated in \Fref{fig:pixel_scale} for a median rms WFE of 350nm, $10^7$ photons per image and 20 Zernike modes. We notice a mild degradation with increasing pixel scale (for pixel scale $>0.5\lambda/D$), and very similar results for small pixel scales ($<0.5\lambda/D$).
 Above the Nyquist limit (pixel scale $<0.5\lambda/D$), the information loss due to the field-of-view cropping is negligible at this level of aberrations and flux. One can expect that in a more aberrated regime, this may start to have a noticeable impact.
Below the Nyquist limit (pixel scale $>0.5\lambda/D$), we notice a loss of accuracy of a factor two. That this degradation is not more severe may be related to the limited number of Zernike modes used here, which produce extended signatures in the focal plane. With an increased spatial frequency content and higher level of aberrations, the PSFs would break into many speckles, which is expected to lead to a more serious degradation of the performance.

\section{Discussion}\label{sec:discussion}

\subsection{Comparison to Gerchberg-Saxton phase retrieval}

\begin{figure*}
\begin{subfigure}[b]{\columnwidth}
  \centering
  \includegraphics[width=5.cm]{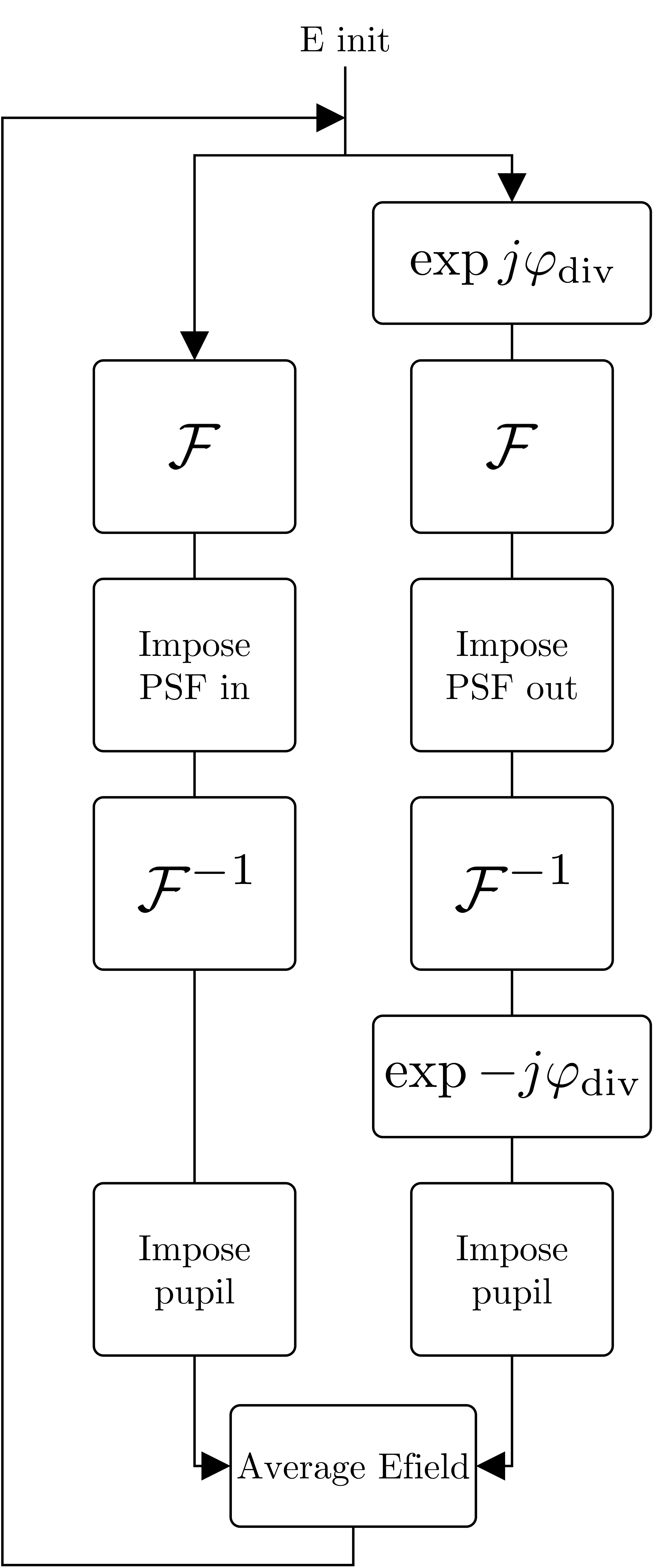}
  \vspace{0.7cm}
  \caption{}
  \label{}
\end{subfigure}
\begin{subfigure}[b]{\columnwidth}
  \centering
  \includegraphics[width=1.0\linewidth]{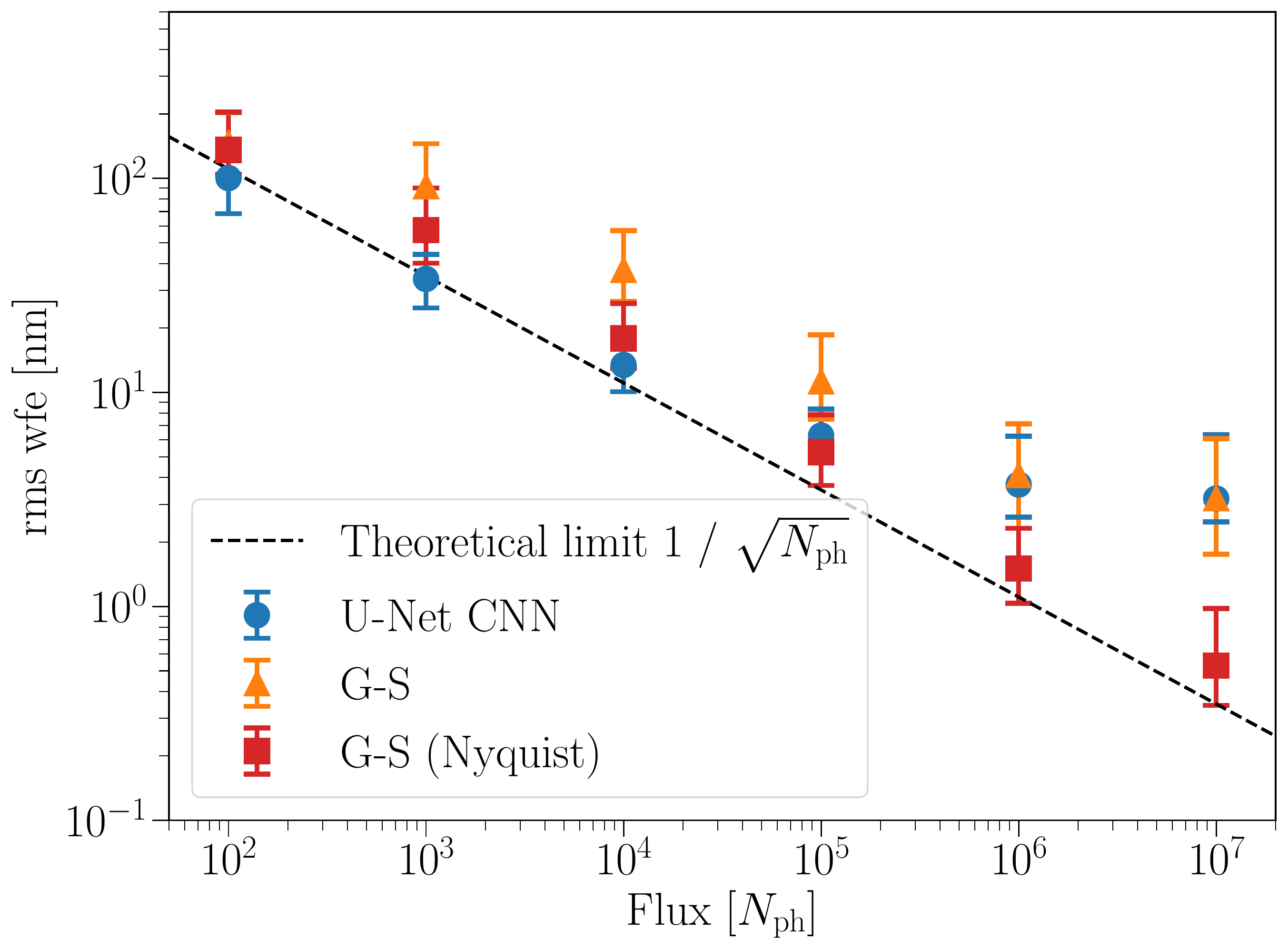}
 \includegraphics[width=1.0\linewidth]{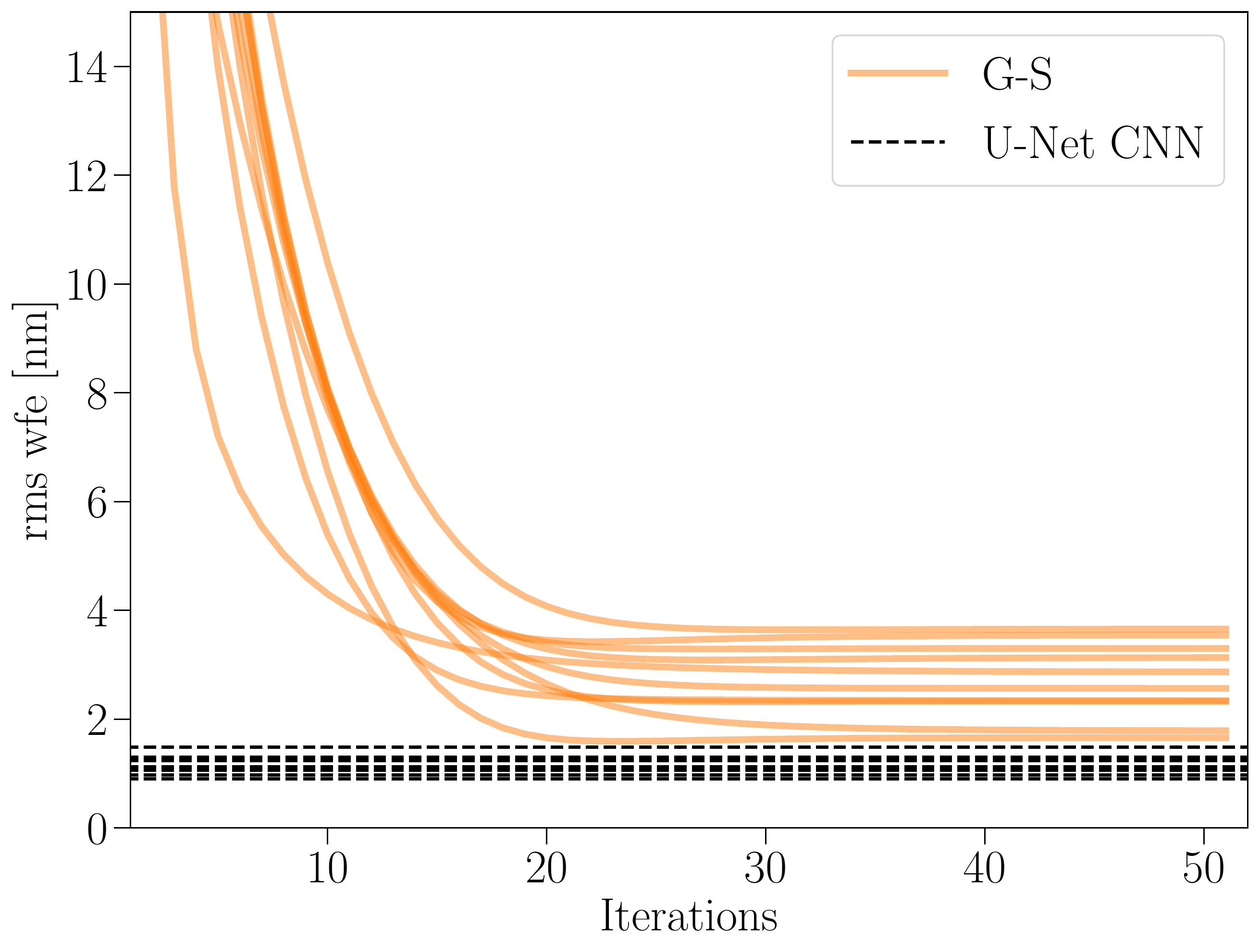}
    \caption{}
  \label{}
\end{subfigure}
\caption{Comparison of an iterative algorithm to the CNN model. (Left) Implemented G-S algorithm for the phase retrieval using two images, one with a phase diversity.
(Top right) Residual rms WFE as a function of photon number per image for the CNN model and for the GS algorithm for two different pixel sampling of the PSFs: in orange, the same sampling used with the CNN, in red, Nyquist sampling. (Bottom right) Illustration of the GS convergence for 10 different evaluations with the rms WFE as a function of iteration, and for a sampling of 0.22$\lambda/D$ per pixels. It takes approximately 20 iterations for the GS algorithm to converge.}
\label{fig:GS}
\end{figure*}

While CNNs can solve the capture range problem and provide an initial estimate for gradient-based optimizers  in the case of large aberrations \citep{Paine:18}, it is also interesting to see how it competes with ``classical" approaches in a lower aberration regime such as explored in the present paper.
A detailed comparison with the most efficient algorithms that may exist is outside the scope of this paper. Rather we compare the CNN models with a standard iterative phase retrieval algorithm to illustrate where the CNN may be superior. The Gerchberg-Saxton iterative algorithm \citep{Gerchberg:72,Fienup:82} is relatively simple to implement, widely used, and can be easily adapted to a specific application.
To exploit our two images, one in-focus and one defocussed, we implement an algorithm that uses multiple images in parallel \citep{Milster:20}.
The algorithm is depicted in \fref{fig:GS} (Left). Since the phase diversity $\varphi_{\rm div}$ is known, it can be appropriately added or removed at different steps of the algorithm. At the end of each iteration, once the phase diversity is removed, the pupil plane electric fields are averaged and the output is used for the next iteration. We compared this parallel approach to a serial one \citep[e.g.][]{Guyon:10,Milster:20} and found it to be superior.

Since the phase is inferred from the complex exponent, the phase output from the iterative algorithm needs to be unwrapped. With phase maps of about 1 radian rms, phase wrapping can occur (the phase can locally be larger than $\pi$) and we need to take it into consideration. Phase unwrapping can be challenging and it is interesting to note that it can also be solved by CNNs \citep{Wang:19}.
To avoid unnecessary complications we assess the performance of the iterative algorithm by analysing the phase residual directly, calculated by $\angle \exp{(i (\varphi - \hat{\varphi}))}$, which we  expect to be in the range of $[-\pi, \pi]$.

We reproduce the analysis in \sref{sec:limit} and evaluate both approaches at different signal levels. For the iterative algorithm, we consider two different pixel scales: 0.5$\lambda/D$ per pixel (Nyquist sampling) and $0.22\lambda/D$ per pixel (same sampling as used for the CNN, see \sref{sec:dataset}).
With Nyquist sampled PSFs, the GS algorithm provides an accuracy close to the theoretical limit over the full range of photon levels, see \Fref{fig:GS} (Top Right). It surpasses the CNN model at high flux, where it does not reach a plateau. 
With the finer sampling, the GS algorithm becomes less accurate at all signal levels and reaches a plateau at high flux similar to the CNN.
We can only suspect that the cropping of the PSFs and the way we impose the amplitude in the image-plane have a detrimental effect on the GS algorithm. While the plateau seen with the CNN is apparently not due to pixel sampling (see \sref{sec:pxscale}), it may be alleviated by a larger training dataset (see \sref{sec:size}).

We also monitor the convergence of the GS algorithm, see \Fref{fig:GS} (bottom right). We can see that the iterative algorithm needs approximately 20 iterations to converge where the CNN performs a similar or superior inference in just one iteration. While we do not quantify the computational gain, this likely translates into a speed advantage for the CNN. For example, \cite{Paine:18} indicated a gain of a factor $\sim 80$ when comparing to nonlinear optimization methods.\\

Finally we also explore the performance obtained with a gradient-based optimizer. More specifically, similarly to \cite{Peng:2020}, we implement the image formation in PyTorch and use its autodifferentiation capabilities to optimize the objective function\footnote{We use here the mean square error of the amplitudes in the focal plane.} using the same variant of stochastic gradient descent, \ie the Adam optimizer. Our results,  not illustrated here for the sake of conciseness, suggest a similar performance to the GS algorithm but requiring a higher number of iteration. Hence, in the context of our paper, gradient-based optimization does not appear to be outperforming the GS algorithm (whose performance are already close to the theoretical limit), nor does it require fewer iterations. A thorough exploration and comparaison of iterative methods is however outside the scope of this paper.

\subsection{Numerical considerations}

In this last section we discuss two numerical aspects relevant for practical applications~: the computational cost associated with the CNN models and the influence of the training dataset size on performance.

\subsubsection{Computational cost}
Computational cost is often measured by the number of floating-point operations (FLOPs) or  the number of multiply-accumulated operations (MACs).
The number of MAC, as well as the number of parameters and the memory size of the two CNN models are  estimated with the package THOP\footnote{ See also \texttt{https://github.com/Lyken17/pytorch-OpCounter} -- , which only considers the number of multiplication operations, to evaluate the computational costs of our two CNN architectures.} and are given in \tref{tab:computation}.
 The number of FLOPs is about twice the number of MACs.

\begin{table}
  \centering
  \caption{Computational cost of the two CNN architectures for 100 Zernike modes.}
  \label{tab:computation}
  \begin{tabular}{lccc}
    \hline
    Architectures & Number of parameters & MACs     & Model size \\
                  &      ($10^6$)        & ($10^9$)      & (MB) \\
    \hline
    ResNet-50 &    23.71  &  4.11  & 91 \\
    U-Net     &    13.40  &  7.77 & 52\\
    \hline
  \end{tabular}
\end{table}

In the case of NCPA, where a correction is expected at best on a timescale of a second, the computational cost given in \tref{tab:computation} is perfectly acceptable with at most 8 - 16 GFLOPs per second.
In contrast, for an AO system typically running at 1kHz, about 8 - 16 TFLOPs per second would be required. Considering that a good GPU RTX 2080Ti provides $>$13 TFLOPs per second in single precision, it can be considered as a feasible approach from the computational power standpoint. Its practical implementation, ranging from an appropriate software implementation to keeping latencies to a minimal level and synchronising the estimation time from multiple GPUs, if needed, might however not be trivial.

However, the numbers given in \tref{tab:computation} should only be considered as upper bounds. Indeed, the deployement of CNN-based models on a real setup could use compression and acceleration techniques to reduce the memory and computational cost \citep[e.g.,][]{Cheng:20}. Compression and speed-up ratios of about 5 to 10 have actually been reported for ResNet and similar architectures \citep[e.g.,][]{Wang:17}.

\subsubsection{Influence of training dataset size}\label{sec:size}

\begin{figure}
  \includegraphics[width=\linewidth]{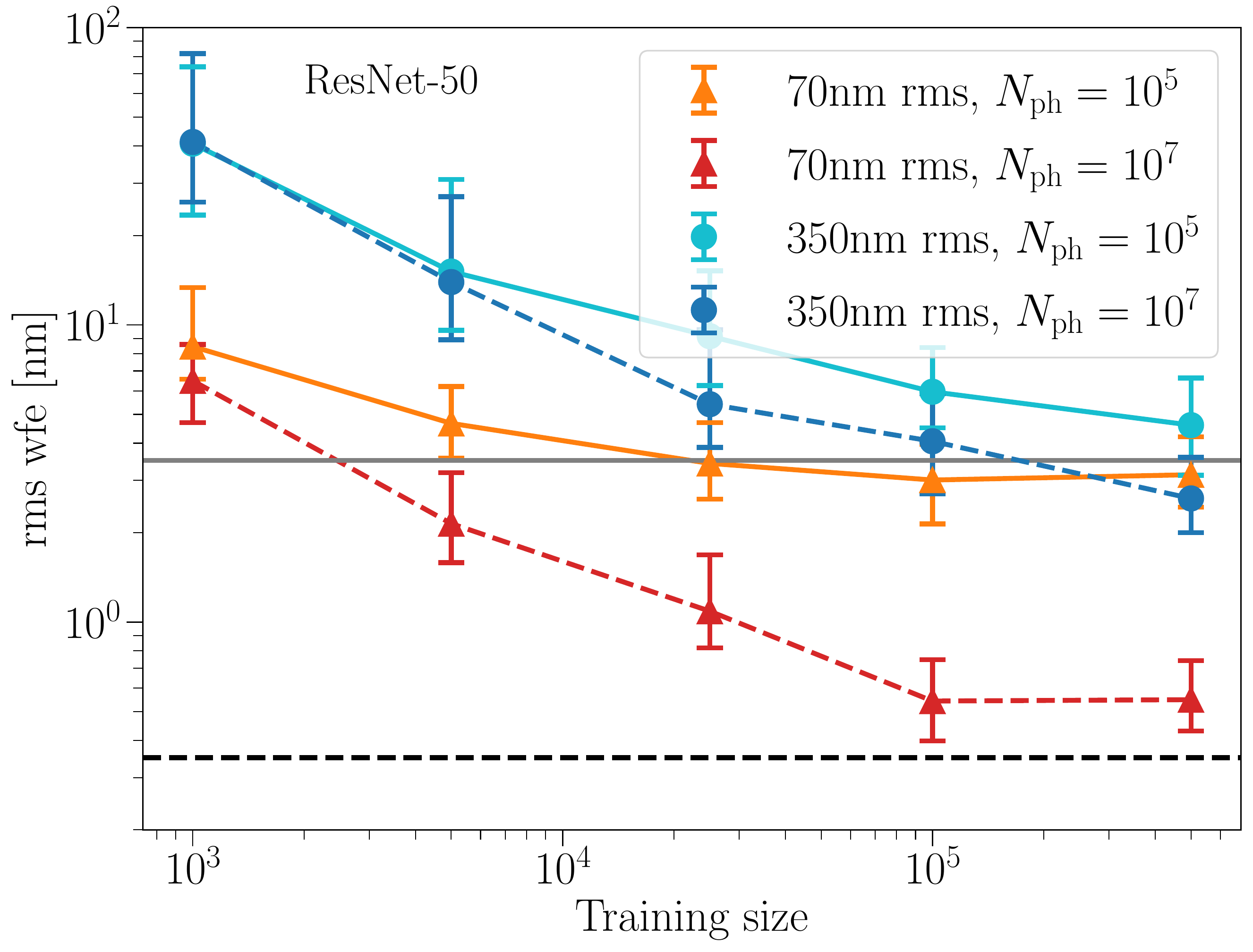}
  \includegraphics[width=\linewidth]{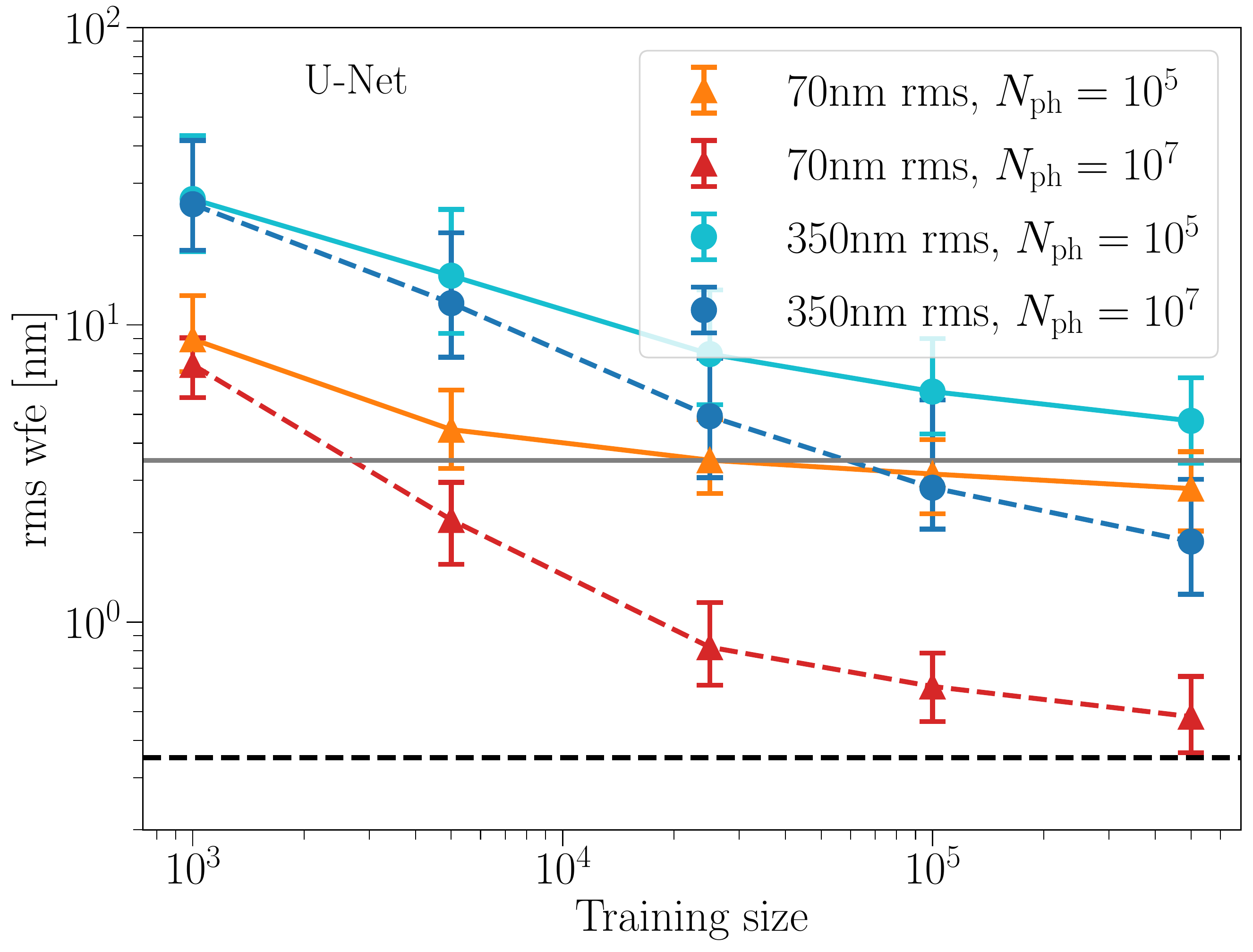}
  \caption{Residual rms WFE as a function of training dataset size for different flux levels and amplitude of the input aberrations. The phase errors are distributed over 20 Zernike modes. The horizontal lines (solid and dashed) represent the theoretical limit for a flux of $ 10^5$  and $10^7$ photons respectively. (Top) ResNet-50. (Bottom) U-Net. }
  \label{fig:training_size}
\end{figure}

To try and understand what may limit the CNN models accuracy, we study the delivered accuracy as a function of the training dataset size.
We compare  two levels of aberrations (distributed over 20 Zernike modes) and two levels of photon noise for the two architectures, ResNet-50 and U-Net. The training dataset sizes range from 1,000 to 500,000 entries, while in the previous sections we used 100,000 entries. The results are illustrated in \fref{fig:training_size}.

In the low aberration (70nm rms) and low flux ($10^5$ photons/image) regime, the accuracy quickly converges with increasing dataset size and requires only $>$5,000 entries. Increasing the flux by two orders of magnitude ($10^7$ photons/image), the SNR is boosted by a factor ten and finer details can be picked up during training. The dataset then needs to be larger with $>100,000$ entries for the training to fully converge and to reach the sub-nm theoretical floor.
For aberrations five times larger (350nm rms), the larger parameter space requires a substantially larger dataset with $\gtrsim 500,000$ entries to attain the theoretical limit in the low flux case.
Finally, at higher flux, we do not reach the theoretical limit, which would require a much larger dataset, \ie $>>500,000$ entries.
Comparing ResNet-50 and U-Net, we can observe that both architectures reach very similar performance.

\begin{figure}
  \centering
  \includegraphics[width=\linewidth]{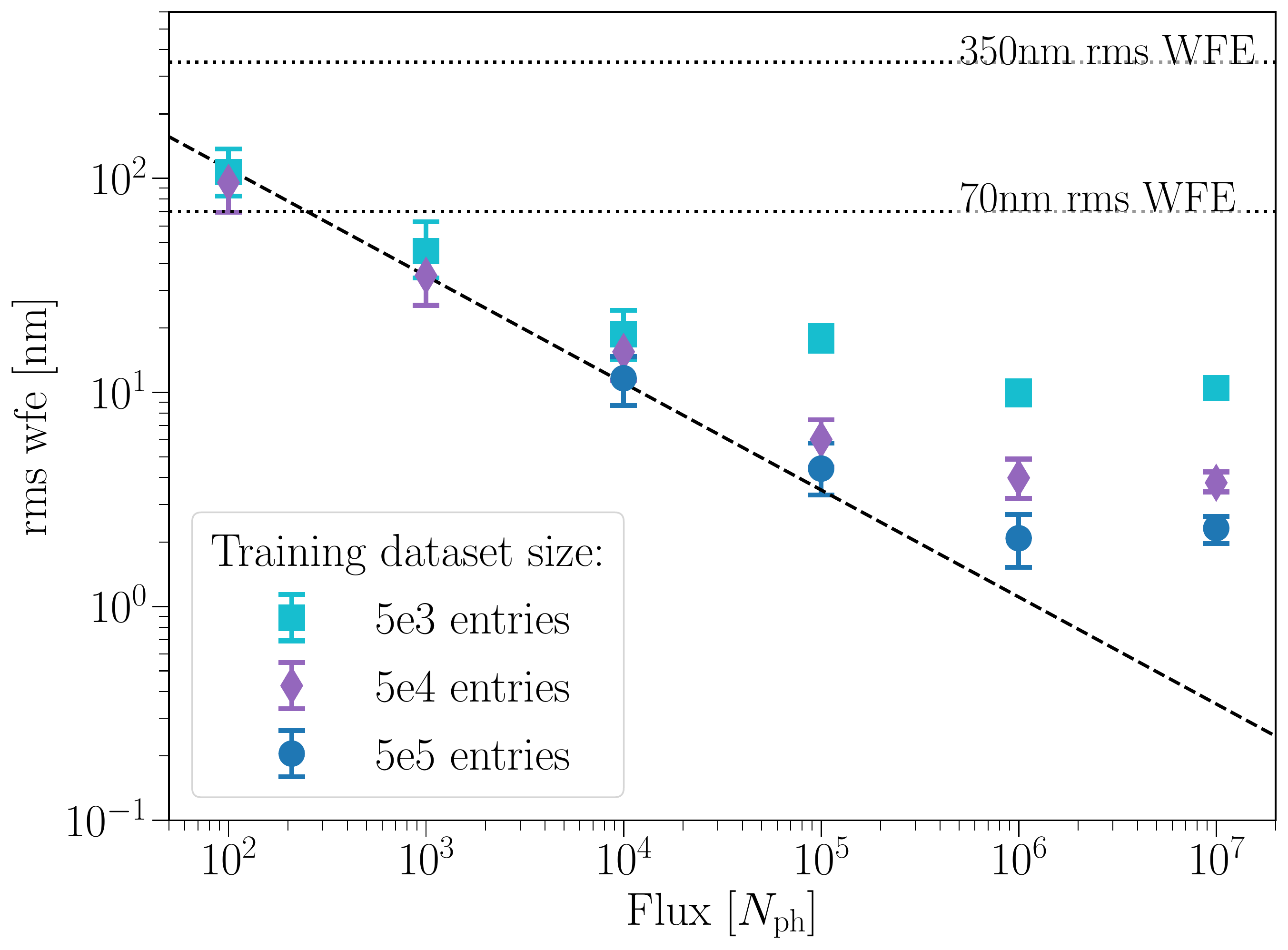}
  \caption{Residual rms WFE as a function of flux level for three training dataset sizes: 5000, 50000, and 500000 entries respectively, using the U-Net model. Increasing the dataset size by a factor ten reduces the model error by a factor of two approximately, at high flux levels.}
  \label{fig:training_size_FP}
\end{figure}

To emphasize the effect of flux, which directly relates to the amount of extractable information, we plot in \fref{fig:training_size_FP} the rms WFE as a function of photon level for three different training set sizes.
The interpretration is relatively straightforward: in each case the accuracy is close to the theoretical limit until it reaches a plateau. The level of this plateau depends on the training dataset size, where a 10-fold increase leads to approximately a factor two improvement of the rms WFE.

The analysis performed here illustrates the need for large training sets when the problem parameter space  increase  if one wants to reach the finest accuracy, i.e. the photon noise limit. For NCPA correction, the range of dataset sizes explored here is presumably large enough. For adaptive optics which has likely a larger parameter space (larger range of aberration amplitudes and larger number of modes to be controlled) but possibly a lower accuracy requirement, one may trade-off accuracy with dataset size for practical reasons, in particular if the training set is composed of experimental data.

\section{Conclusions}\label{sec:conclusion}

In this paper, we explored the use of deep convolutional neural networks (CNNs) to perform image-based wavefront sensing. We focussed on low level of aberrations (0.1 - 1rad rms WFE) and a limited number of spatial frequencies (20 - 100 Zernike modes). Those parameters are characteristic of NCPA measurements on  large ground-based telescopes (8 - 40m, e.g. VLT or ELT). Our simulations suggest that the CNN models are able to leverage the high sensitivity of focal plane wavefront sensing 
over a broad range of signal levels.
In terms of dynamic range, we have demonstrated successful correction for aberration levels up to $\lambda/6$ rms WFE in one iteration, and $\gtrsim\lambda/2$ in 5 to 10 iterations.
The models are robust under reasonable flux changes, with a mild departure from the photon noise limit with changing SNR level.
The prediction accuracy of the trained models is however affected by unknown disturbances such as higher-order aberrations, and the training strategy should be adapted accordingly.

 The type of architecture, and in particular the choice of approach between the Zernike coefficients (ResNet) versus the direct phase map (U-Net) estimations, has a negligible impact in our experiments \citep[contrary to, e.g.,][]{Guo:19}.
In fact, we have used ResNet-50 and U-Net interchangeably. U-Net does have a slight edge in terms of generalization power, as illustrated in \fref{fig:ho}, but this advantage is too marginal to justify alone a preference for this architecture.
When compared to an iterative phase retrieval algorithm, the CNN models display similar and often superior accuracies in just one iteration, while the iterative algorithm requires 10 to 20 iterations to converge. Hence, in addition to a close-to-optimum estimation accuracy, CNN models are expected to be faster.

While using CNN-based FPWFS for NCPA measurement seems readily applicable, its utilization for adaptive optics appears more challenging for an equivalent telescope diameter.
Indeed, adaptive optics calls for the sensing of a larger number of modes  (at least a factor 10) and larger aberration levels (a factor $\sim$10 in the bootstrapping phase of the AO closed-loop) compared to NCPA measurement,  hence the dimensionality is largely increased compared to NCPA measurement. Finally, AO also requires sub-ms inference speed potentially constraining the CNN architecture and its implementation.
In the prospect of real application, an encouraging result  of our simulation  is that CNNs can be applied iteratively, in closed-loop, to reduce the wavefront error to a low level. In particular the wavefront error stays at a stable low level, close to the expected theoretical limit, and is able to converge in just a few iterations for initial aberration levels well beyond its training range.

Real-life data is however more complex than our simulated images, including their finite wavelength range, the different detector noises, the residual atmospheric perturbations, imperfect optical alignment, etc.
Some of those effects can be anticipated and simulated, such as the polychromaticity, which will wash out part of the high spatial frequency information, but some others may not be.
While the instrument model can be improved to generate more realistic labelled datasets, ultimately experimental data are needed for the training of the CNN models. How to best exploit a limited experimental dataset is thus one of the key challenge for future applications.
Another key aspect in the context of NCPA measurements is to maintain a 100\% science duty cycle despite the  phase diversity required to lift any ambiguity in the FPWFS measurement. An interesting avenue in that context is to leverage the diversity introduced by the changing atmosphere using for example Long Short Term Memory (LSTM) networks to exploit its temporal structure.

\section*{Acknowledgements}
This research used different Python packages and libraries:
Pytorch \citep{paszke2017automatic}, Astropy \citep{astropy:2013}, AOtools \citep{Townson:19}, PROPER \citep{proper}, Numpy \citep{harris2020array}, Matplotlib \citep{Hunter:2007}.
The research was supported by the European Research Council (ERC) under the European Union’s Horizon 2020 research and innovation program (grant agreement No 819155), and by the Wallonia-Brussels Federation (grant for Concerted Research Actions).

\section*{Data Availability}

The data underlying this article will be shared on reasonable request to the corresponding author.

\bibliographystyle{mnras}
\bibliography{ML_WFS}


\appendix


\bsp	
\label{lastpage}
\end{document}